\documentclass{cernrep}

\usepackage{fancyhdr}
\fancyhfoffset{4 mm}
\fancypagestyle{ARTTITLE}{%
\fancyhf{} % clear all header and footer fields
\lhead{\small{Proceedings of the CERN--Accelerator--School course: \it{Introduction to Accelerator Physics}}}
\lfoot{Available online at \url{https://cas.web.cern.ch/previous-schools}}
\rfoot{\thepage\hspace*{3mm}}

}

\usepackage[colorlinks=true, linkcolor=black, citecolor=black, filecolor=black, urlcolor=blue]{hyperref}

\begin{document}
\title{Superconducting Magnets}
\author{Gijs de Rijk}
\institute{CERN, Genève, Switzerland}

\begin{abstract}
Superconductivity allows to construct and operate magnets at field values beyond 2 Tesla, the practical limitation of normal-conducting magnets exploiting ferro-magnetism. The field of superconducting magnets is dominated by the~field generated in the coil. The stored energy and the electromagnetic forces generated by the coil are the main challenges to be overcome in the design of these magnets.

For further reading you may consult the following books: \cite{bib:Wilson}, \cite{bib:Mess},\cite{bib:Iwasa},\cite{bib:ROXIE} or the proceedings of two specialized CAS courses: \cite{bib:Brandt} and \cite{bib:Turner}.
\end{abstract}


\maketitle
\thispagestyle{ARTTITLE}

% SECTION
\section{Introduction: magnetic fields and basic magnets}
% SUBSECTION
\subsection{Magnetostatics}
A magnet at constant field can be described by the Maxwell equations with the time dependent terms set to zero, the case of magnetostatics. Let's have a closer look at the three equations that describe magnetostatics (in the differential form). \newline 
Gauss law of magnetism: \hspace{5.9cm} \begin{math} div \overrightarrow{B} = 0 \end{math} \hfill(always holds). \newline 
Ampère's law with no time dependencies:  \hspace{3.35cm}  \begin{math} rot \overrightarrow{H} = \overrightarrow{J} \end{math}  \hfill(magnetostatics) \newline 
Relation between the magnetic field $\overrightarrow{H}$ and the flux density  $\overrightarrow{B}$:   \hspace{0.1cm}   \begin{math} rot \overrightarrow{B} =  \mu_{0} \mu_{r} \overrightarrow{H} \end{math}       \hfill(linear materials)

% SUBSECTION
\subsection{Coil dominated magnets}

\begin{figure}[ht!]
\begin{center}
\includegraphics[height=2.5cm]{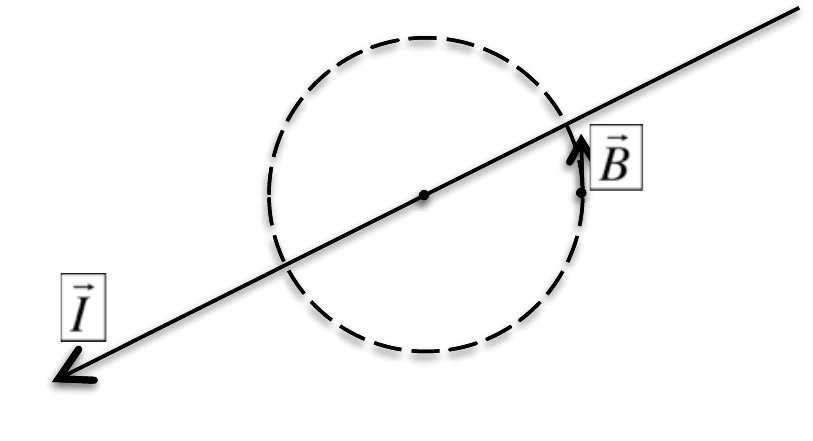} \hspace{ 2cm}
\includegraphics[height=2.5cm]{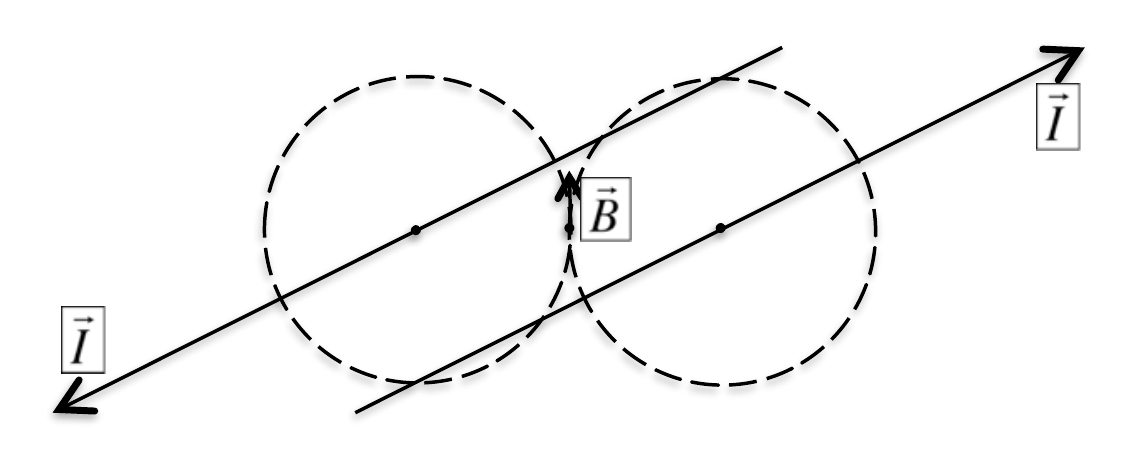}
\caption{left: infinite wire with circular field line; right: two infinite wires with opposite currents}
\label{fig:current-wires}
\end{center}
\end{figure}

From Ampère's law with no time dependencies (integral form), $ \oint_c \overrightarrow{H}\cdot \overrightarrow{{\rm d}l}  =  \mu_{0} I $ we can derive the law of Biot and Savart for the field generated by a current carrying line (see Fig.~\ref{fig:current-wires} left):  $ \overrightarrow{B} = \frac{\mu_{0}I}{2 \pi r} \cdot \hat{\varphi}$ with $\hat{\varphi}$ the direction vector as the tangent of a circle.

\begin{figure}[ht!]
\begin{center}
\includegraphics[width=5.5cm]{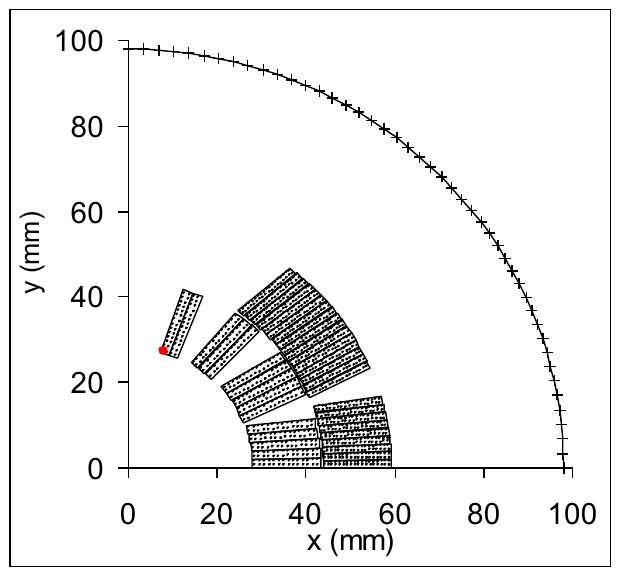}
\caption{LHC dipole: a quarter of the coil of one aperture}
\label{fig:LHCdipXsection}
\end{center}
\end{figure}

If we want to make a  $B=8 \UT$ magnet with just two infinitely long thin wires placed at a distance of $50 \Umm$ in air (see Fig.~\ref{fig:current-wires} right) we need $I = 5\cdot10^{2}\UA$. We see that to reach high fields ($ \geq 4 \UT$) we need very large currents, moreover in such configurations the field quality will be poor.
Let us compare this to the LHC main dipole magnets. In Figure~\ref{fig:LHCdipXsection} a quarter of the coil of one aperture of an LHC dipole is shown. The full coil has 80 turns that with a current of $I = 9.48\cdot10^{5}\UA$ produces a field of $8.34\UT$. This means that the average current density in the coil area of the magnet is $\approx 300\UA/ \UmmZ^{2}$. We can see that to get high fields we need very high currents through small surfaces.

% SUBSECTION
\subsection{Magnetic field quality: Multipole description}
The beam in a particle accelerator is confined in a beam pipe around which the bending and focusing magnets have to be arranged.  From beam dynamics calculations one can derive that the field quality is very demanding; for the dipole magnets that provide the bending of the beam, the typical required field homogeneity is: $ \frac{\Delta B_{z}}{|B|} \leq 10^{-4} $. 
The field quality in accelerator magnets is expressed and measured in a~multipole expansion:
\begin{equation}
B_{y}+{\rm i}B_{x} = 10^{-4}B_{1} \sum_{n=1}^{\infty}(b_{n}+{\rm i}a_{n}) \left ( \frac{x+{\rm i}y}{R_{\rm ref}} \right )^{n-1} \; \mathrm{with} \ b_{n}, a_{n} \leq \mathrm{few \ units}.  \label{eq:a8}
\end{equation}
With: $z=x+iy$; $B_{x}$ and $B_{y}$ the flux density components in the $x$ and $y$ direction; $R_{ref}$ the radius of the reference circle; $B_{1}$ the dipole field component on the reference circle; $b_{n}$ and $a_{n}$ the normal and skew $n^{th}$ multipole component. The "wanted" $b_{m}$ or $a_{m}$ is equal to one. \\
In a circular accelerator, where the beam makes multiple passed, one typically demands that for all the other components than the wanted component: \\ $a_{n}, b_{n} \leq 10^{-4}$.

% SUBSECTION
\subsection{Magnetic length}
\begin{figure}[ht!]
\begin{center}
\includegraphics[height=5.5cm]{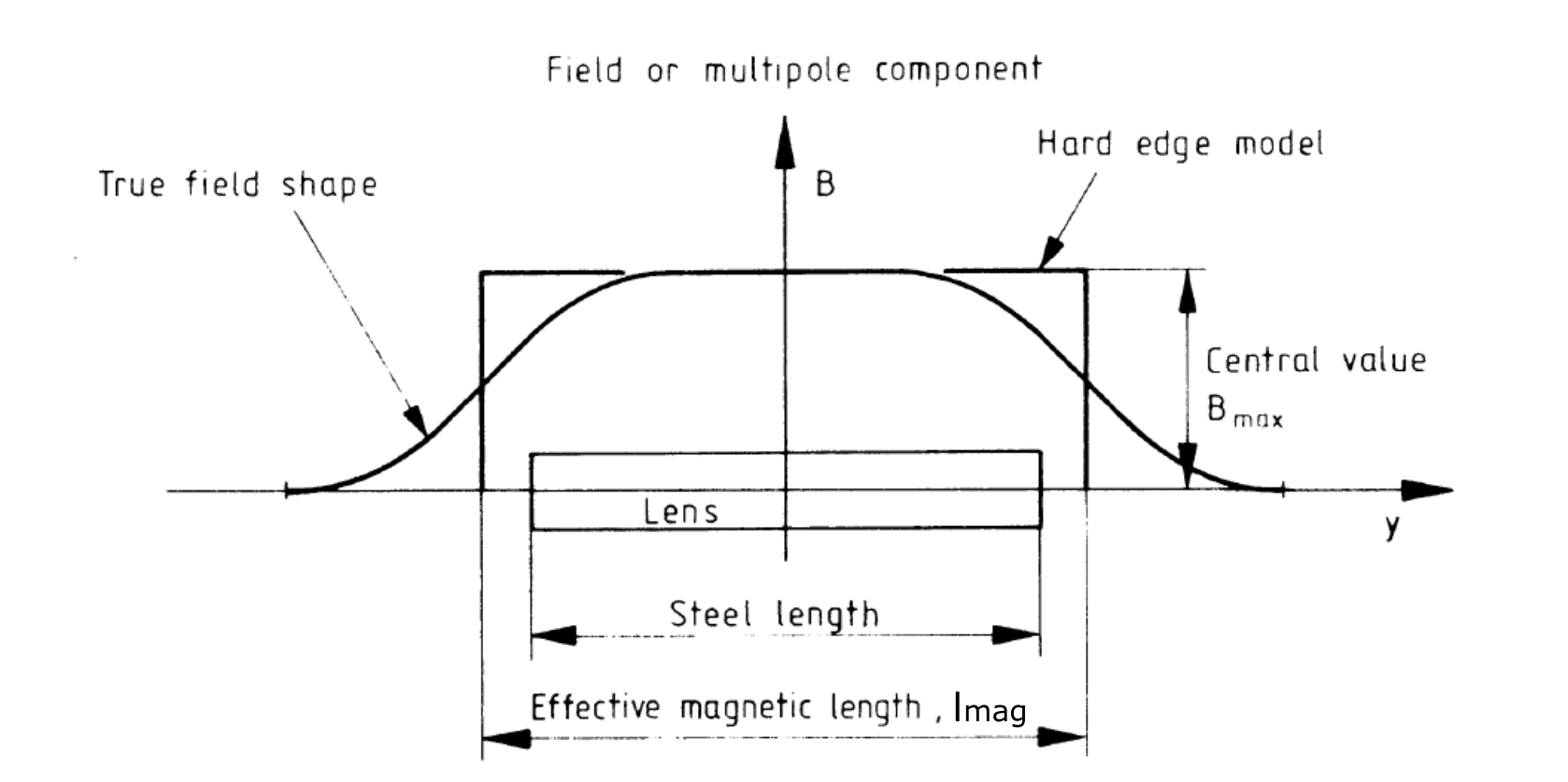}
\caption{Shape of the magnetic field in the longitudinal dimension (named y in this plot)}
\label{fig:axialfield}
\end{center}
\end{figure}

In the longitudinal dimension the typical shape of a magnetic field can be seen in Fig.~\ref{fig:axialfield} .
We can define the magnetic length $l_{mag}$ from: $ l_{mag}B_{0} = \int_{-\infty}^{+\infty}B(z)dz$ with $B_{0}$ the field at the centre of the magnet $z=0$. \\
The magnetic length $L_{mag}$ of superconducting magnets, that have a cylindrical yoke around the coils, is adjustable by varying the length of the yoke and the coils are often sticking out on both sides of the yoke. There is hence no easy rule-of-thumb for $L_{mag}$ for superconducting magnets.

% SUBSECTION
\subsection{Generating a perfect dipole field}
\begin{figure}[ht!]
\begin{center}
\includegraphics[height=4.5cm]{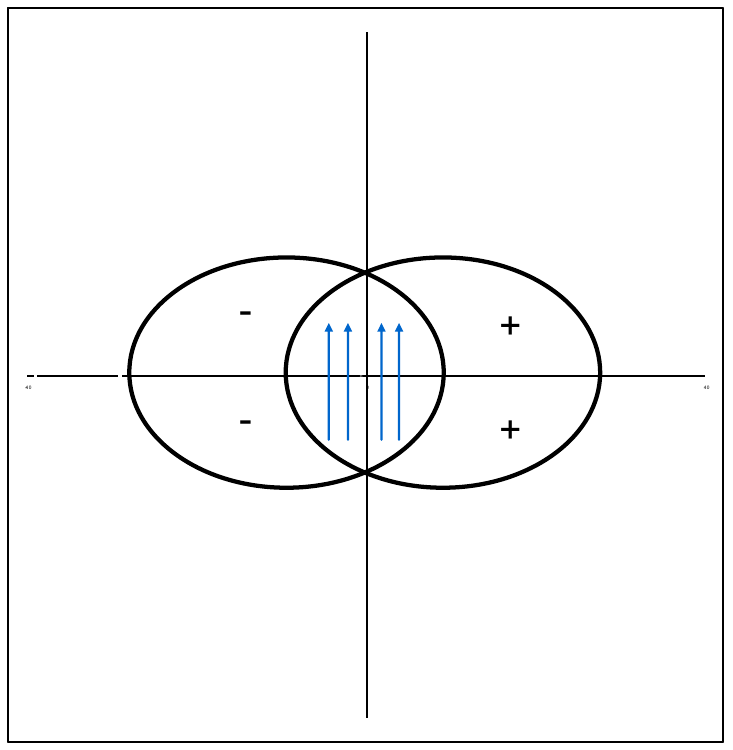} \hspace{ 0.5cm}
\includegraphics[height=4.5cm]{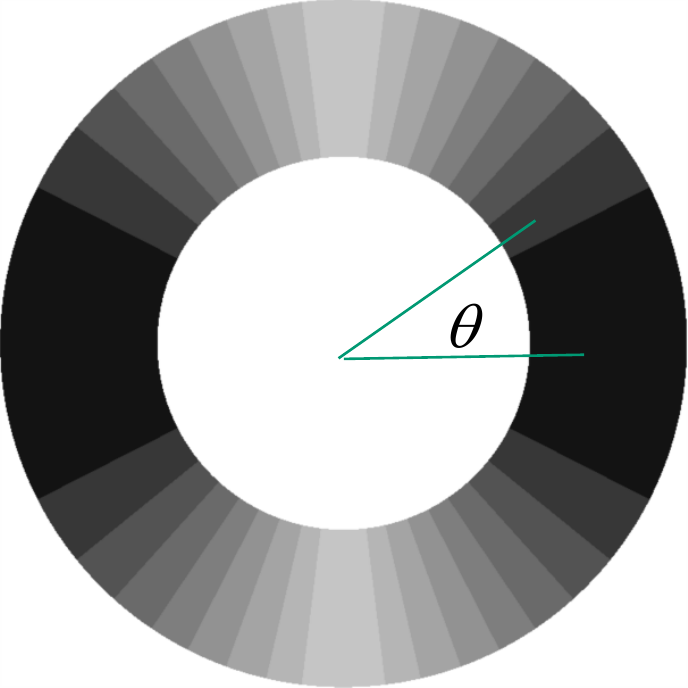}  \hspace{ 0.5cm}
\includegraphics[height=4.5cm]{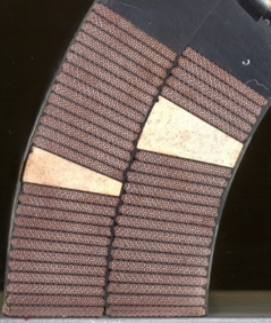}
\caption{left: Two intercepting ellipses with uniform opposite currents   ; middle: $cos(\Theta)$ current distribution; right:  Flat rectangular cables in a quarter of a $cos(\Theta)$ coil }
\label{fig:costh}
\end{center}
\end{figure}

Let us consider the case were we want to generate a perfect dipolar field with conductors in a simple geometry. The case of two solid intercepting ellipses (or circle as special case of an ellipse) that are uniformly conducting current in opposite directions generate in the overlap region a perfect dipolar field (Fig.~\ref{fig:costh} left). We can remark that the field region ("magnet aperture") is not circular and that moreover such a geometry is difficult to realise with a flat cable conductor.
A pure dipolar field can also be generated by a thick conductor shell with a $cos\Theta$  current distribution:   $J = J_{0} cos(\Theta)$  (Fig.~\ref{fig:costh} middle). Such a configuration is easier to reproduce with with a flat rectangular (or slightly key-stoned) cable (Fig.~\ref{fig:costh} right).

\begin{figure*}[ht!]
\begin{center}
\includegraphics[width=14.cm]{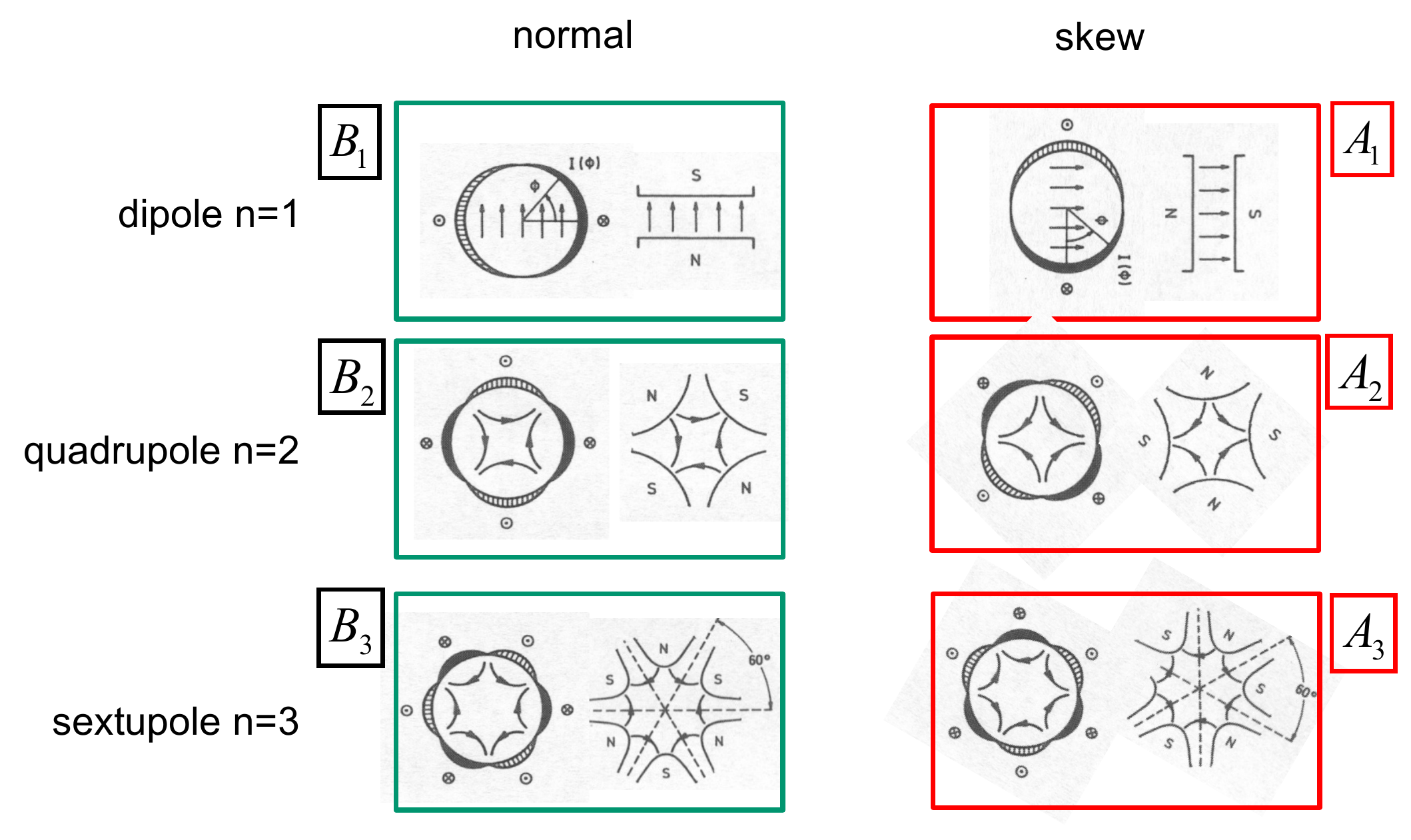}
\caption{Magnets types with $cos\Theta $ coils according to the main field pole they supply up to $n=3$ for both normal and skew poles.}
\label{fig:scmagtypes}
\end{center}
\end{figure*}

% SUBSECTION
%\subsection{Magnetic types with $cos(\Theta)$ coils}
Using the $cos(\Theta)$ coil layout as introduced in the previous section we can construct magnets with any multipolar index $n$. In Figure~\ref{fig:scmagtypes} the magnets types are shown according to the main field pole they supply up to $n=3$ for both normal and skew multipoles. 

% SECTION
\section{State of the art superconducting accelerator magnets}
\begin{figure}[ht!]
\begin{center}
\includegraphics[height=3.5cm]{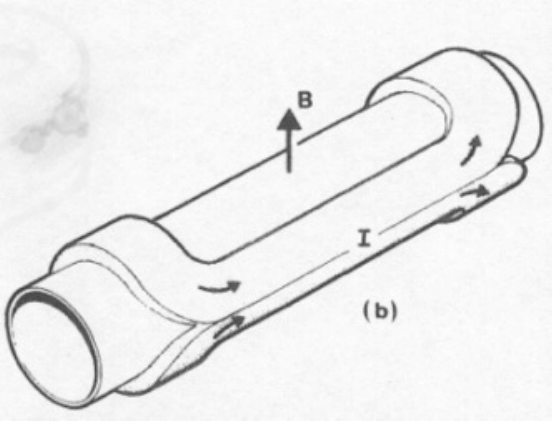} 
\includegraphics[height=3.5cm]{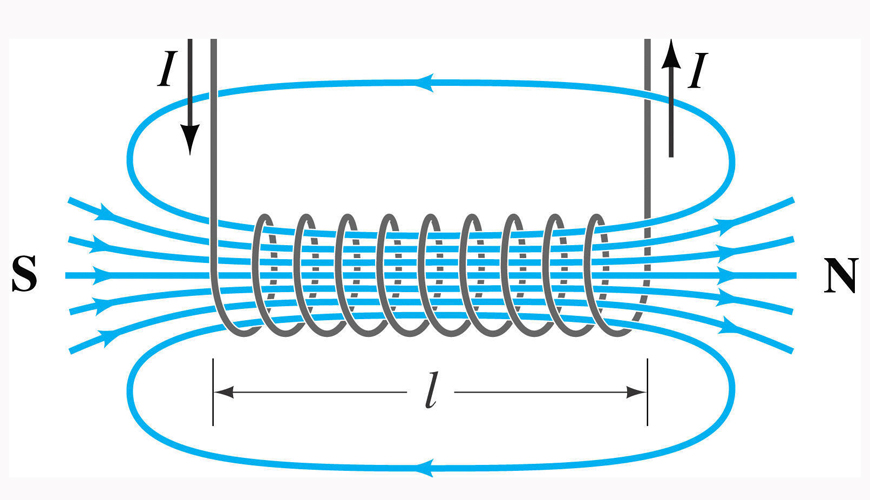} 
\includegraphics[height=3.5cm]{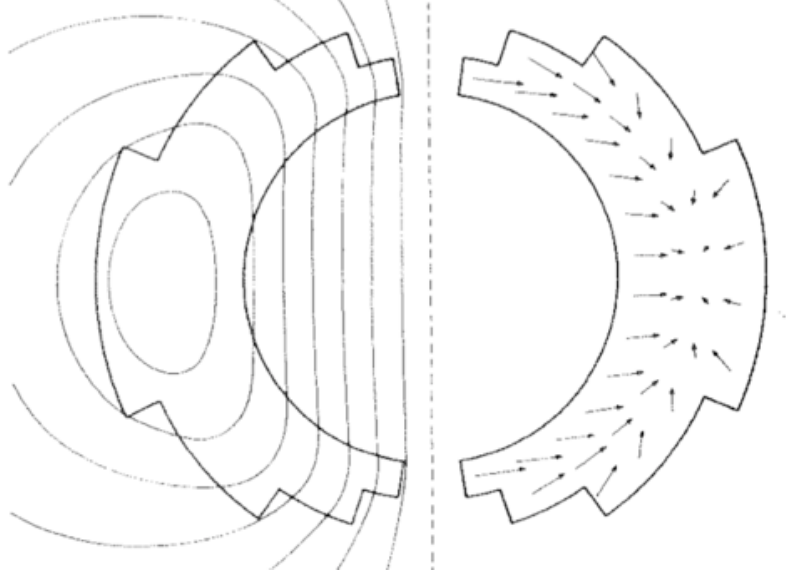} 
\caption{left: Coil with a perpendicular field ; middle: Solenoid coil with an axial field; right: Forces in a $cos(\Theta)$ dipole coil}
\label{fig:accmag}
\end{center}
\end{figure}

In order to bend the beam around in a circular trajectory the field has to be perpendicular to the beam direction. The field layout of an accelerator magnet is thus with the field direction perpendicular to the axial axis of the aperture (Fig.~\ref{fig:accmag} left). The resulting coils are not as efficient as for a solenoid coil (Fig.~\ref{fig:accmag} middle), where the coil completely envelops the field area. If one looks at the Lorentz forces on the coil then in the case of a solenoid they are pointing radially outwards. Such a force can be held efficiently by a surrounding cylinder or simply by the conductor itself that will then be in tension. For a $cos(\Theta)$ coil in a dipole magnet with a vertical field, the force will have two components: one compresses the coil on the~horizontal mid-plane and the other is pointing sideways outwards in the mid-plane (Fig.~\ref{fig:accmag} right). Such forces are more complicated to contain and pose one of the challenges for superconducting accelerator magnet design. 
 
\begin{figure}[ht!]
\begin{center}
\includegraphics[height=5cm]{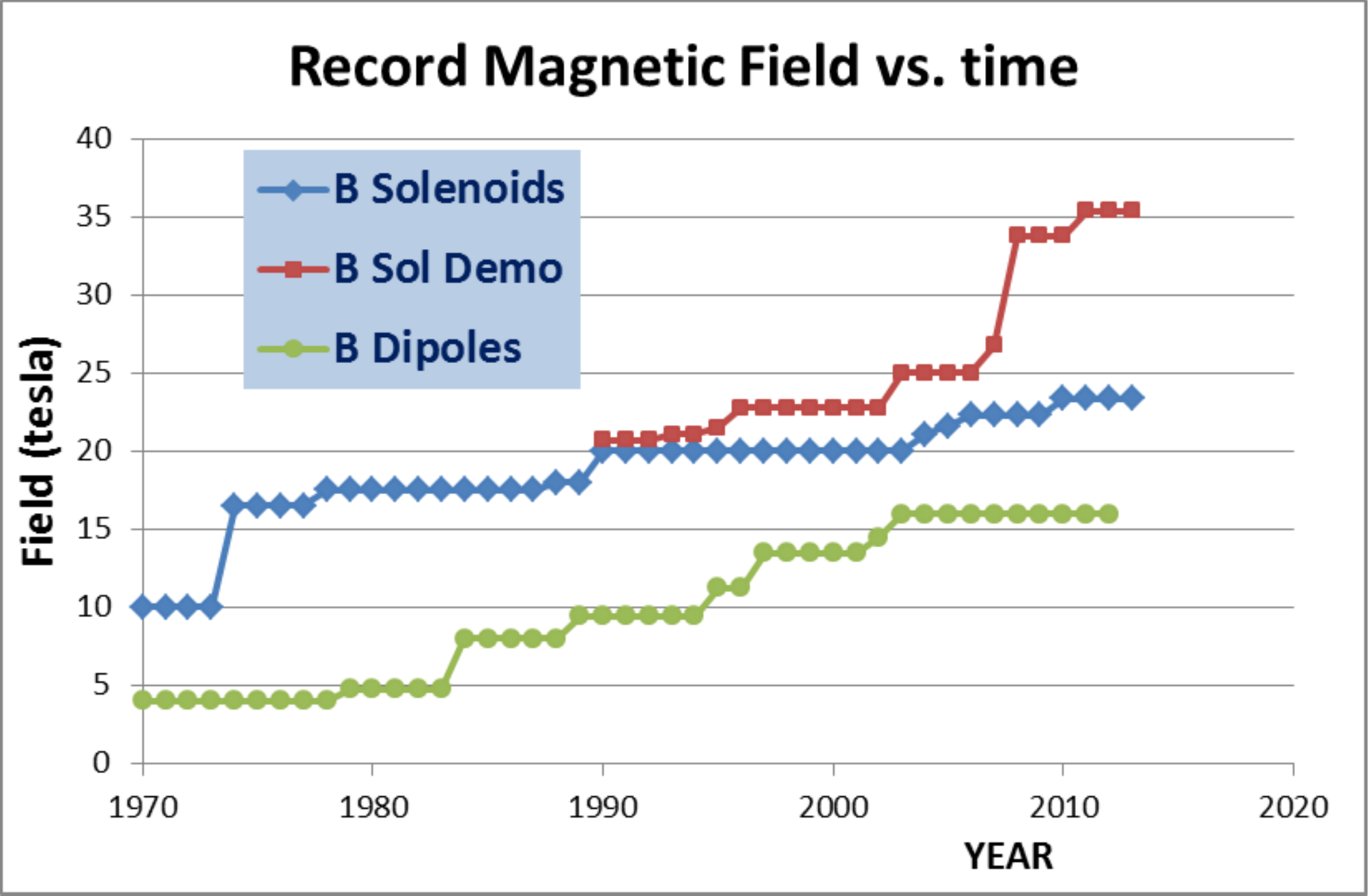} \hspace{0.3cm}
\includegraphics[height=5cm]{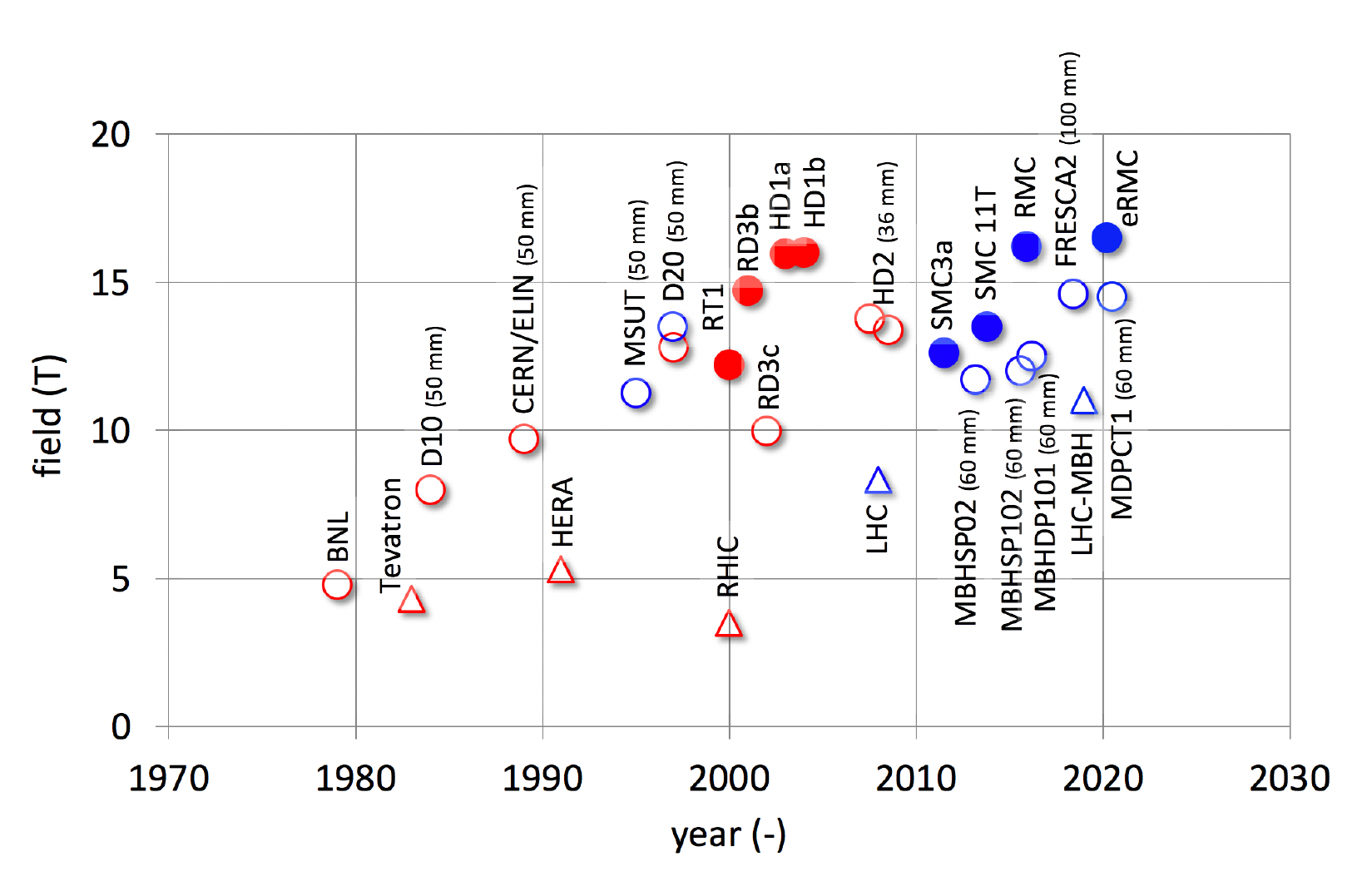} 
\caption{left: Record fields of solenoids and dipoles over time ; right: record fields of superconducting dipole as function of time, open symbols are for magnets with a real aperture, solid symbols are for magnets without a usable aperture (e.g. racetrack coils), rectangulars are for magnets that were used in an operational accelerator.}
\label{fig:fielddev}
\end{center}
\end{figure}

The efficiency of axial field solenoid coils with respect to perpendicular field coils is illustrated in Fig.~\ref{fig:fielddev}. On the left side we can see the record magnetic flux density $B$ achieved as function of time with the two types of coils. There is roughly a factor two difference during the last decades in what was achieved with the two types. In the picture on the right we can see a compilation of flux density values used in accelerators and achieved in development magnets during the last decades. The open symbols are for magnets with a usable aperture, the closed ones for magnets that have essentially no aperture for a beam. Later on we will see that an accelerator magnet can at best run with a $20\% $ margin with respect to its maximum attainable field, in order to be sufficiently reliable in an operational machine. At this moment in time the maximum attainable fields are approaching $16\UT$. 

\begin{figure}[ht!]
\begin{center}
\includegraphics[height=5.5cm]{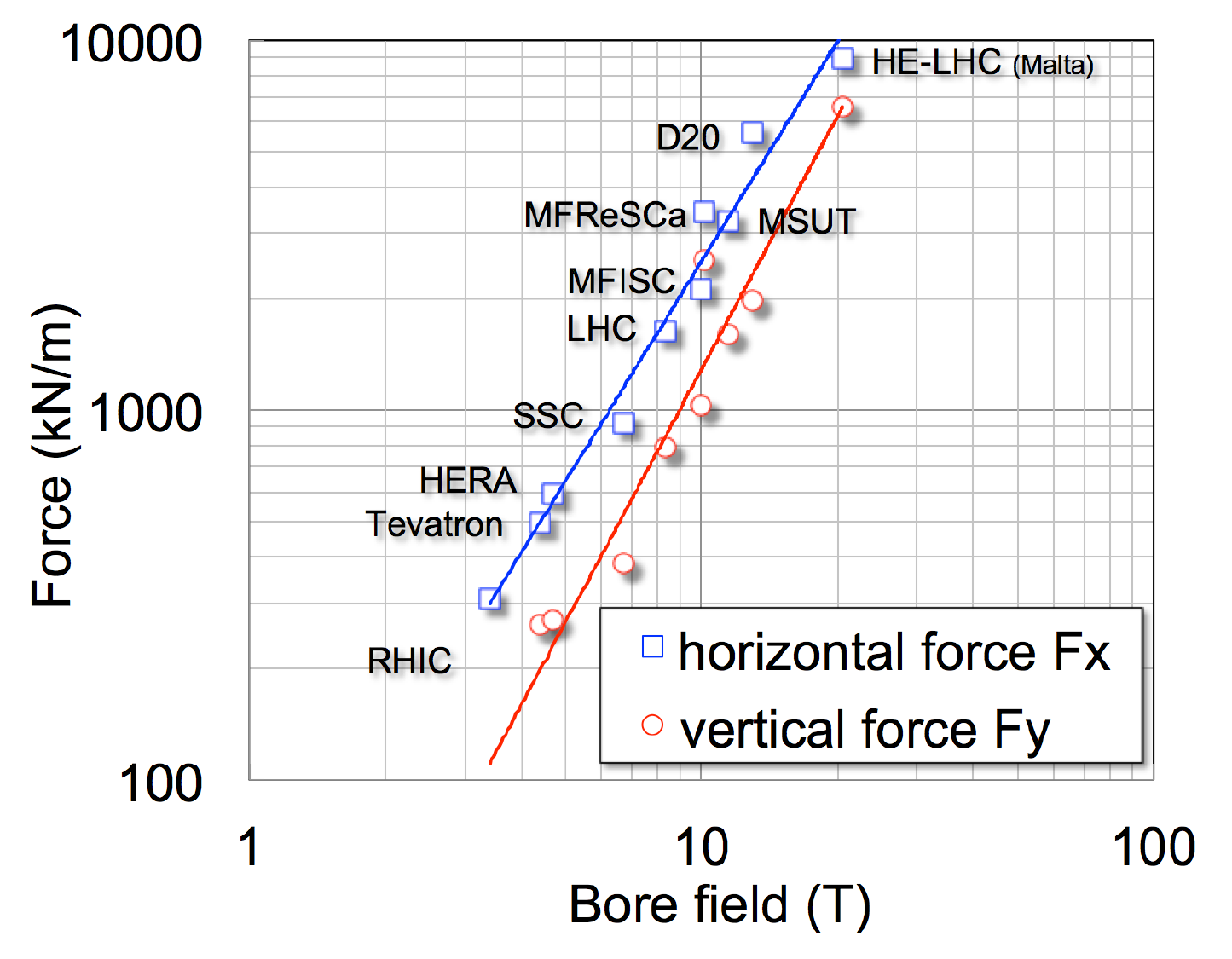} \hspace{ 0.5cm}
\includegraphics[height=5.5cm]{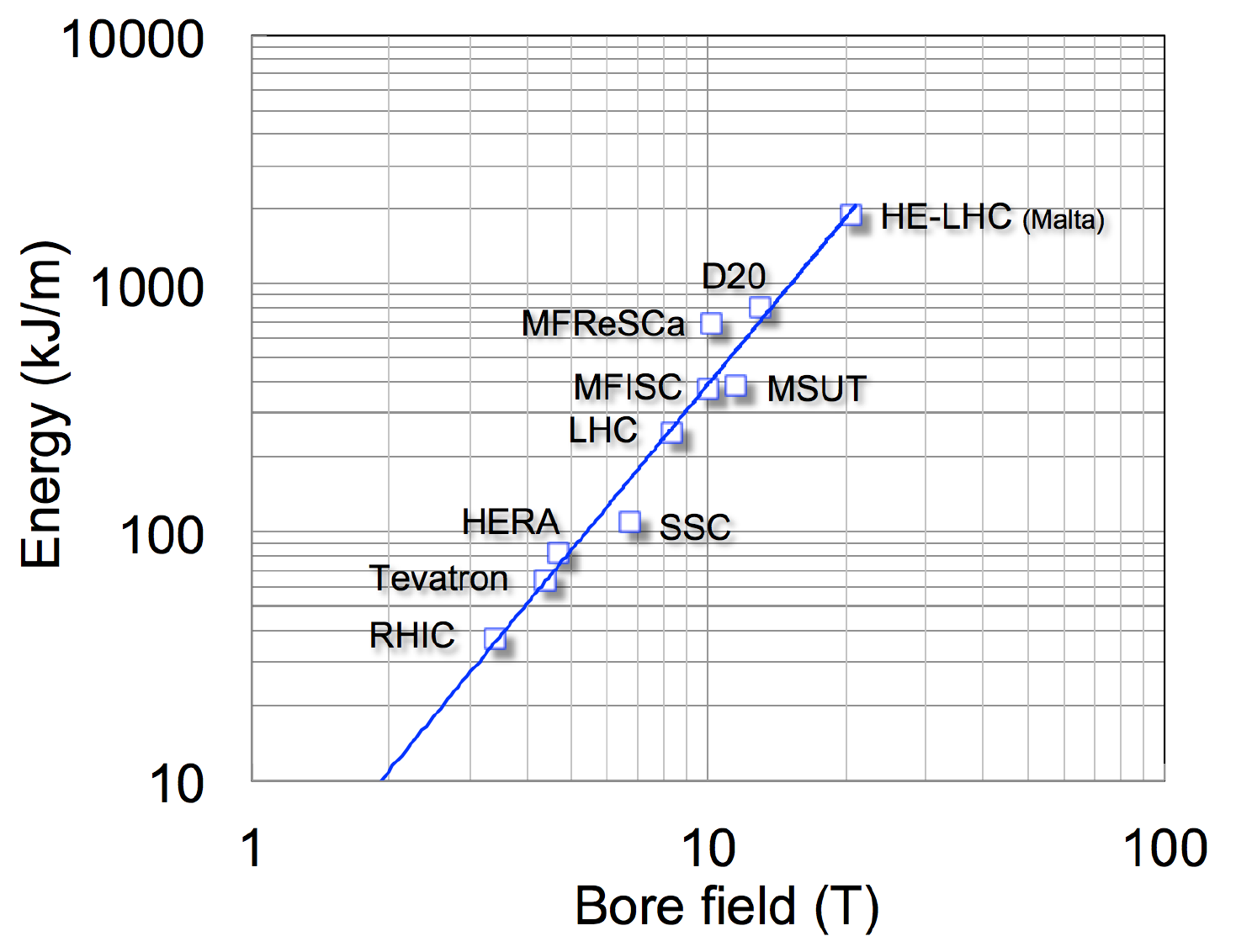}  \hspace{ 0.5cm}
\caption{left: Force in dipole magnets as function of field; right: Magnetic energy in dipole magnets as function of field}
\label{fig:force-energy}
\end{center}
\end{figure}

The two most important physical parameters of magnets that have to be mastered to attain high fields are the electromagnetic force and the electromagnetic stored energy. The electromagnetic forces are the sum of all the Lorentz forces on the individual conductors that carry a current while situated in the field of the other current carrying conductors. In Figure~\ref{fig:force-energy} on the left side on can see the total force as function of the bore field for a quarter of a magnet for a number of magnets that can also represented in the previous figure. The total force scales close to a quadratic curve with the field.  An electromagnetic field represents a stored energy according to  $ E_{mag} = \iiint H B dV$, integrated over the full volume with field. We can see that the total stored energy for magnets also roughly increases quadratically with the~field.

\begin{figure}[tb!]
\begin{center}
\includegraphics[height=9cm]{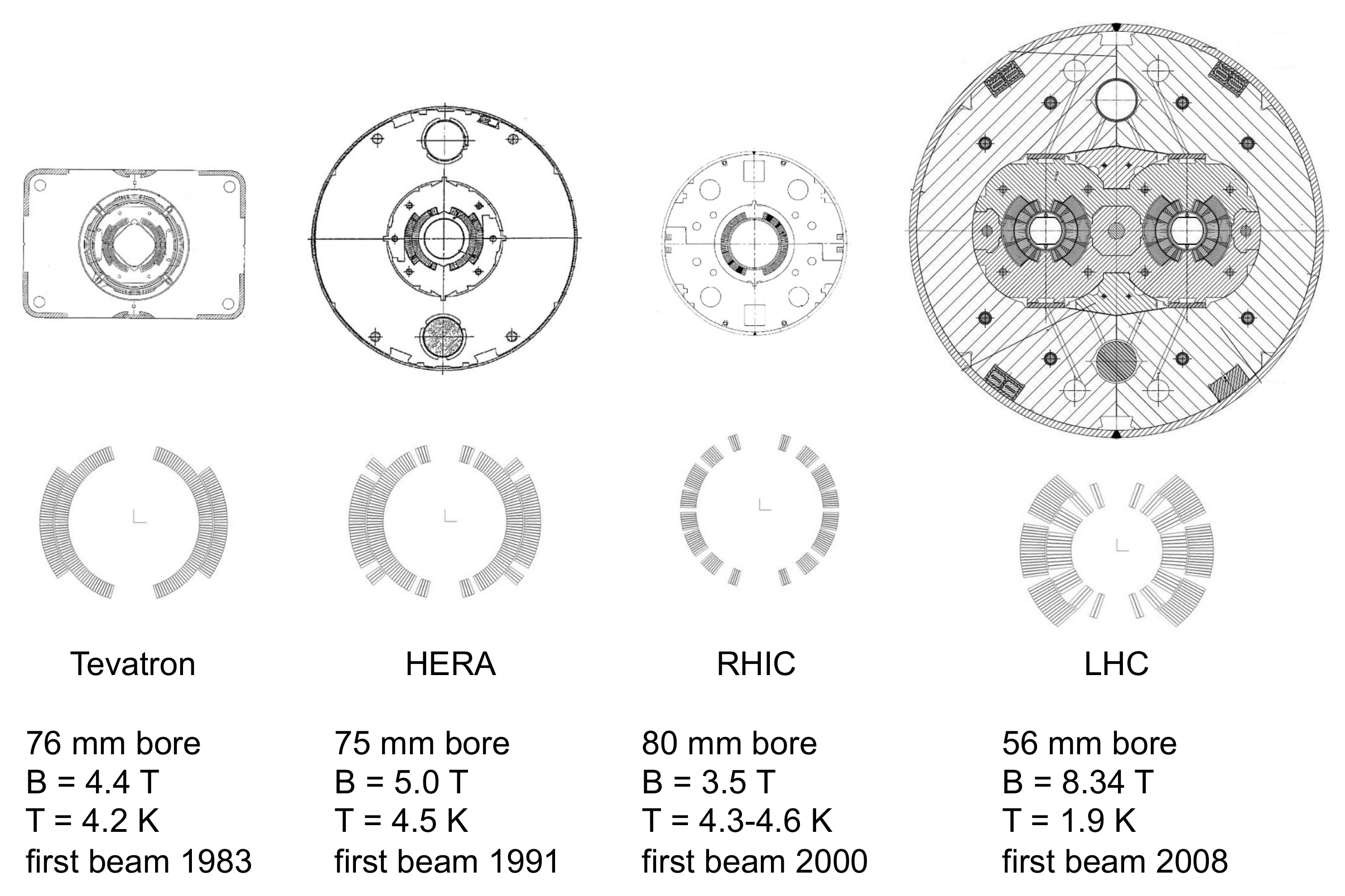} 
\caption{Cross sections of dipole magnets employed in particle accelerators}
\label{fig:runaccdipcomp}
\end{center}
\end{figure}

To make the most efficient use of the space available, to maximise the energy of the accelerator, accelerator magnets are made as long as practically possible. In the Tevatron collider at Ferrmilab the~superconducting dipole magnets were $6\Um$ long, while in the LHC at CERN they are nearly $15\Um$ long. In long magnets the beam trajectory will already bent by a significant angle which brings the beam off-centre in the magnet aperture. To limit the effect of this long magnets are often bent: the LHC dipole magnets are built with a $9.14\Umm$ sagitta.\\
In the end of the 1970s first superconducting magnets started to be employed in accelerators or their beam-lines. In a SPS beam-line at CERN a $2 \Um$ long $B=4.5\UT$ dipole (CESAR) was used next to a~quadrupole (CASTOR). In the ISR collider at CERN low beta quadrupoles were installed to enhance the luminosity; they delivered $40\UT / \Um$ in an aperture of $73\Umm$ diameter. These early magnets all used Nb-Ti conductor in a monolithic rectangular conductor.
In Figure~\ref{fig:runaccdipcomp} magnet cross sections of superconducting magnets that have been, or are being, used in fully superconducting accelerators are shown. The~Tevatron in Fermilab (Chicago, US) was the first fully superconducting synchrotron and started operations in 1983. HERA at DESY (Hamburg, DE) followed in 1991. Using an older magnet design, RHIC at BNL (Brookhaven, US) stared up in 2000 and finally the LHC at CERN (Geneva, CH) in 2008. All these magnets in these four machines employ Nb-Ti conductors in the form of Rutherford cables (details on Rutherford cables in \Sref{sec:Rutherford}) in a $cos(\Theta)$ coil layout.

% SECTION
\section{Superconductors}

\begin{figure}[tb!]
\begin{center}
\includegraphics[height=8cm]{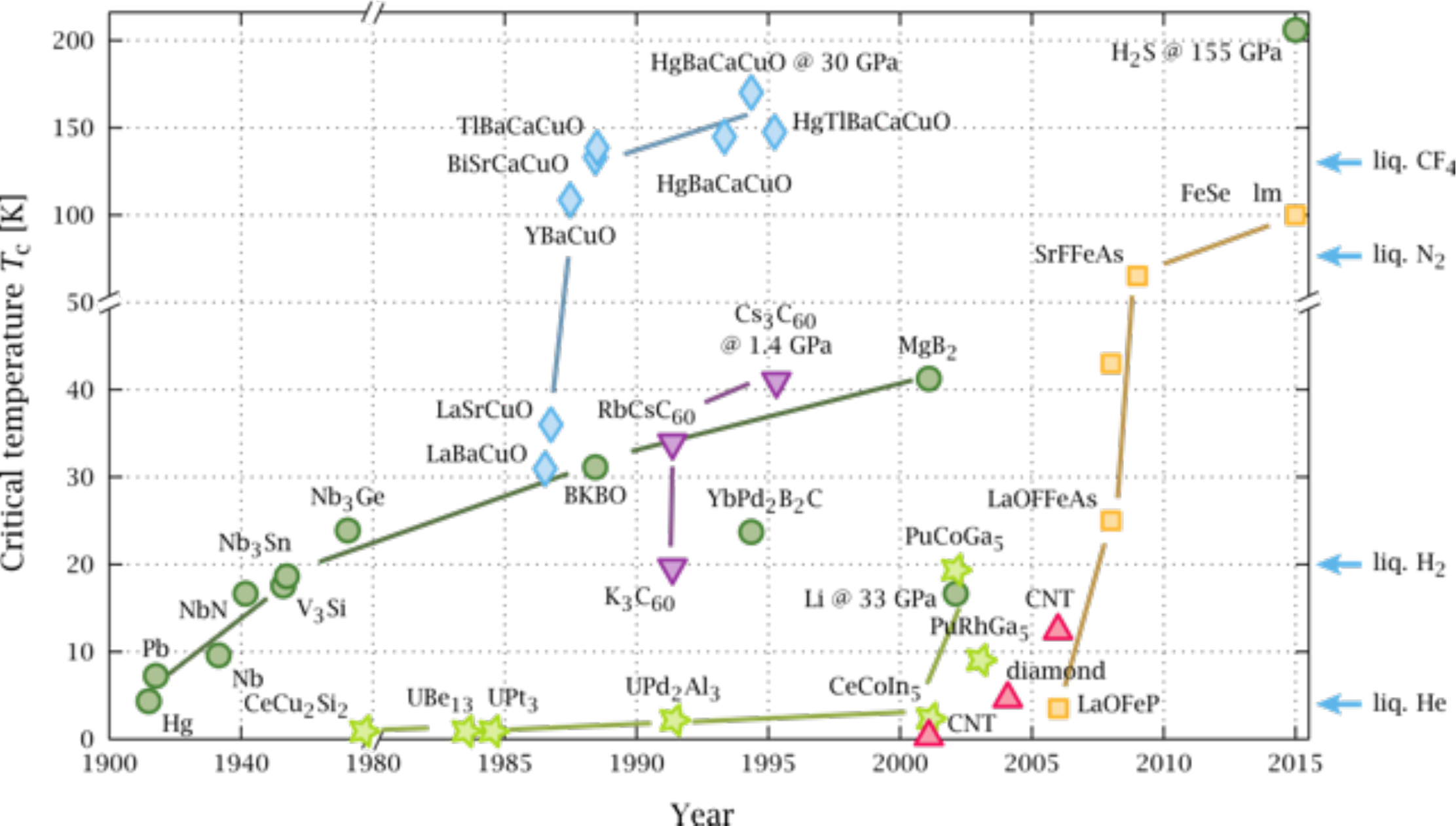} 
\caption{Discovery timeline of superconductors, each type of symbol is for a specific series of substances. e.g.: green circles are for metallic LTS, blue rhombus' are for HTS cuprates, yellow squares are for Iron Based Superconductors (IBS).}
\label{fig:SCdiscov}
\end{center}
\end{figure}
Superconductors form the heart of a superconducting magnet. Superconductivity was discovered in 1911 by Kamerlingh Onnes in Leiden, but it took many decades before technical superconductors came on the~market from which superconducting magnets could reliably be built. Since then regularly new superconducting materials have been discovered. All these materials are metallic and displayed superconducting properties at relatively low temperatures (below $40\UK$), that necessitate cooling with liquid helium or liquid hydrogen and are called Low Temperature Superconductors (LTS). In 1987 a new series of ceramic materials appeared that displayed superconductivity also at higher temperatures, the so-called High Temperature Superconductors (HTS) that come in range of cooling with liquid nitrogen. In Figure~\ref{fig:SCdiscov} a timeline of discovery of superconducting materials is shown.

\begin{figure}[tb!]
\begin{center}
\includegraphics[height=12cm]{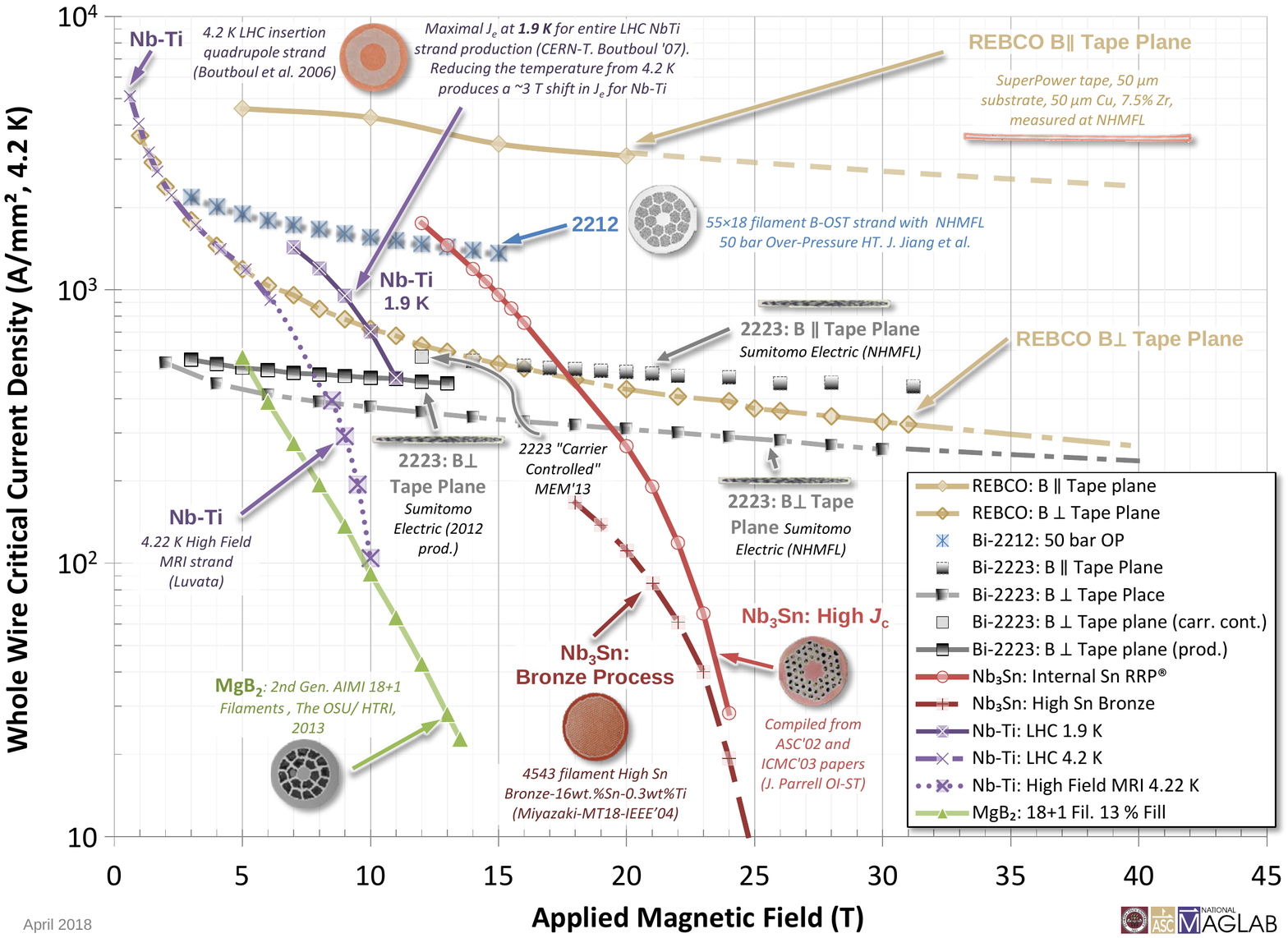} 
\caption{Critical current density averaged over the whole conductor cross section as function of applied magnetic for industrially available superconductors at $4.5\UK$ and $1.9\UK$ (from: \cite{bib:web}) }
\label{fig:jeplot}
\end{center}
\end{figure}

Only a limited number of materials are suitable for magnet construction and have been developed into industrially available conductors. Industrially available materials are shown in Fig.~\ref{fig:jeplot}, where the~critical current density averaged over the~whole conductor cross section is plotted against the applied magnetic field at $4.2\UK$ and $1.9\UK$.

% SUBSECTION
\subsection{Low Temperature Superconductors}
\begin{figure}[ht!]
\begin{center}
\includegraphics[height=4.5cm]{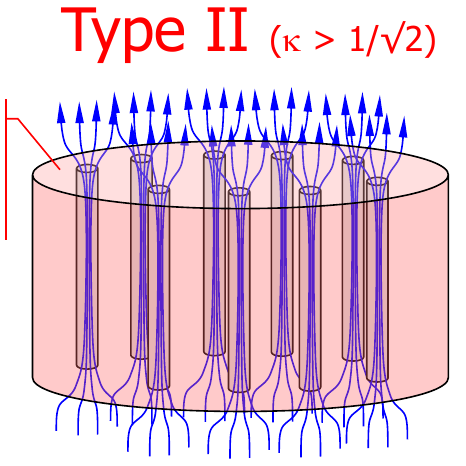} \hspace{1cm}
\includegraphics[height=4.cm]{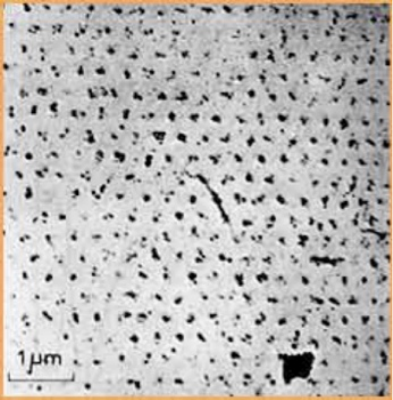} \hspace{1cm}
\includegraphics[height=8cm]{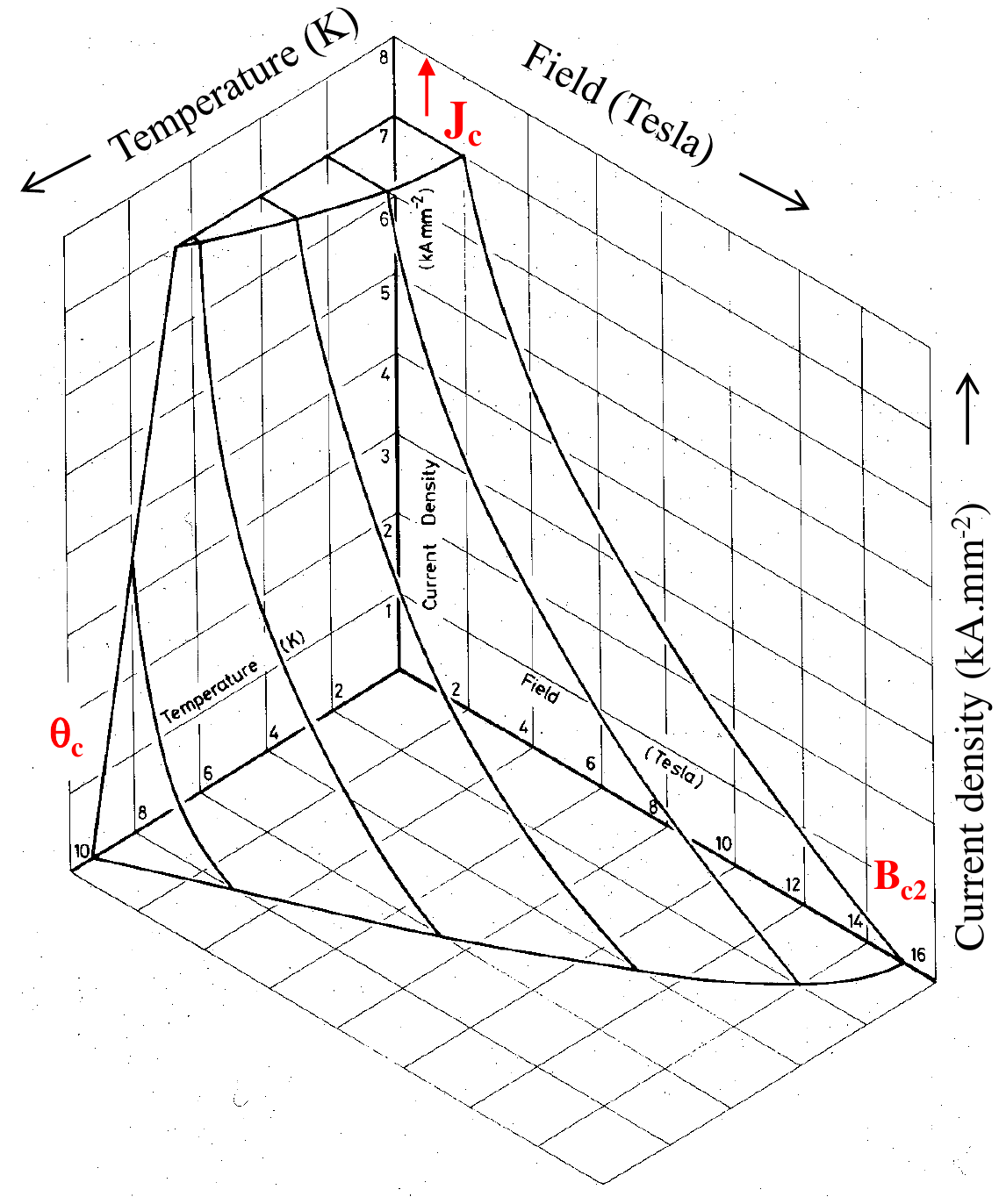} 
\caption{Type-II (LTS) superconductors; top-left: A schematic representation of flux lines in a TypeII superconductor; top-right: Photo of a flux lines exiting from a material visualised with iron particles; bottom: Critical surface plot of a Nb-Ti conductor.}
\label{fig:typeII}
\end{center}
\end{figure}

Of the LTS type of conductors only two materials are being employed: Nb-Ti and Nb$_{3}$Sn. \\
Both LTS superconductors in use are so called TypeII superconductors, where above a critical flux density $B_{c1}$ the field penetrates the superconductor in the form of quantised flux lines. The critical flux density where the whole conductor is fully occupied with flux lines is called $B_{c2}$ and gives the practical field limit for the conductor to be superconducting. In the parameter space Temperature--Flux density--current density the material is superconducting when the state is below a surface, the so-called "critical surface", above this surface the material is normal conducting. The transition through the surface is a very sharp state transition from a zero resistance state to a resistive state. In Figure~\ref{fig:typeII} we can see a sketch of the~flux line penetration in a material and a photo of flux lines exiting from a material visualised with iron particles. We can see in the figure a plot of a critical surface for a Nb-Ti conductor.

\begin{figure}[ht!]
\begin{center}
\includegraphics[height=5cm]{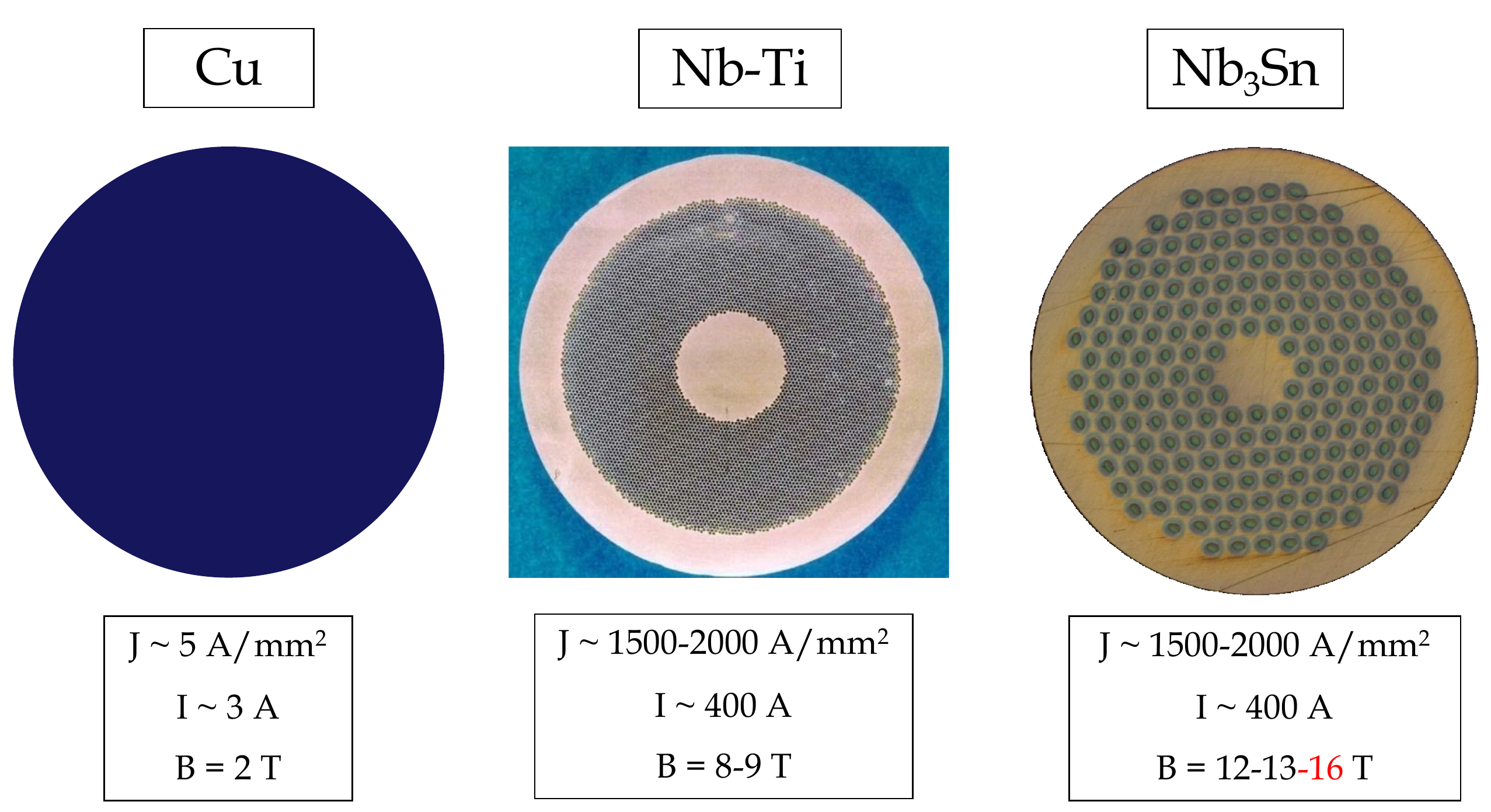} 
\caption{Comparison of 3 conductor wires of the same diameter; Cu, Nb-Ti and Nb$_{3}$Sn}
\label{fig:3wires}
\end{center}
\end{figure}

Practical TypeII conductors are made as thin filaments of the superconducting material inside a~copper matrix that provides stabilisation. In most cases this presents itself as a round wire. In Figure~\ref{fig:3wires} we can see a comparison of three wires (Cu, Nb-Ti and Nb$_{3}$Sn) where we can see the critical current density in the conducting part of the cross section, the total current and the field at which this can be attained. Cu wires can only carry a few $\UAZ/\UmmZ^{2}$ while the superconducting wires have a two orders of magnitude higher current density. Cu has no critical field, while superconductors are limited by this.

\begin{figure}[ht!]
\begin{center}
\includegraphics[height=5cm]{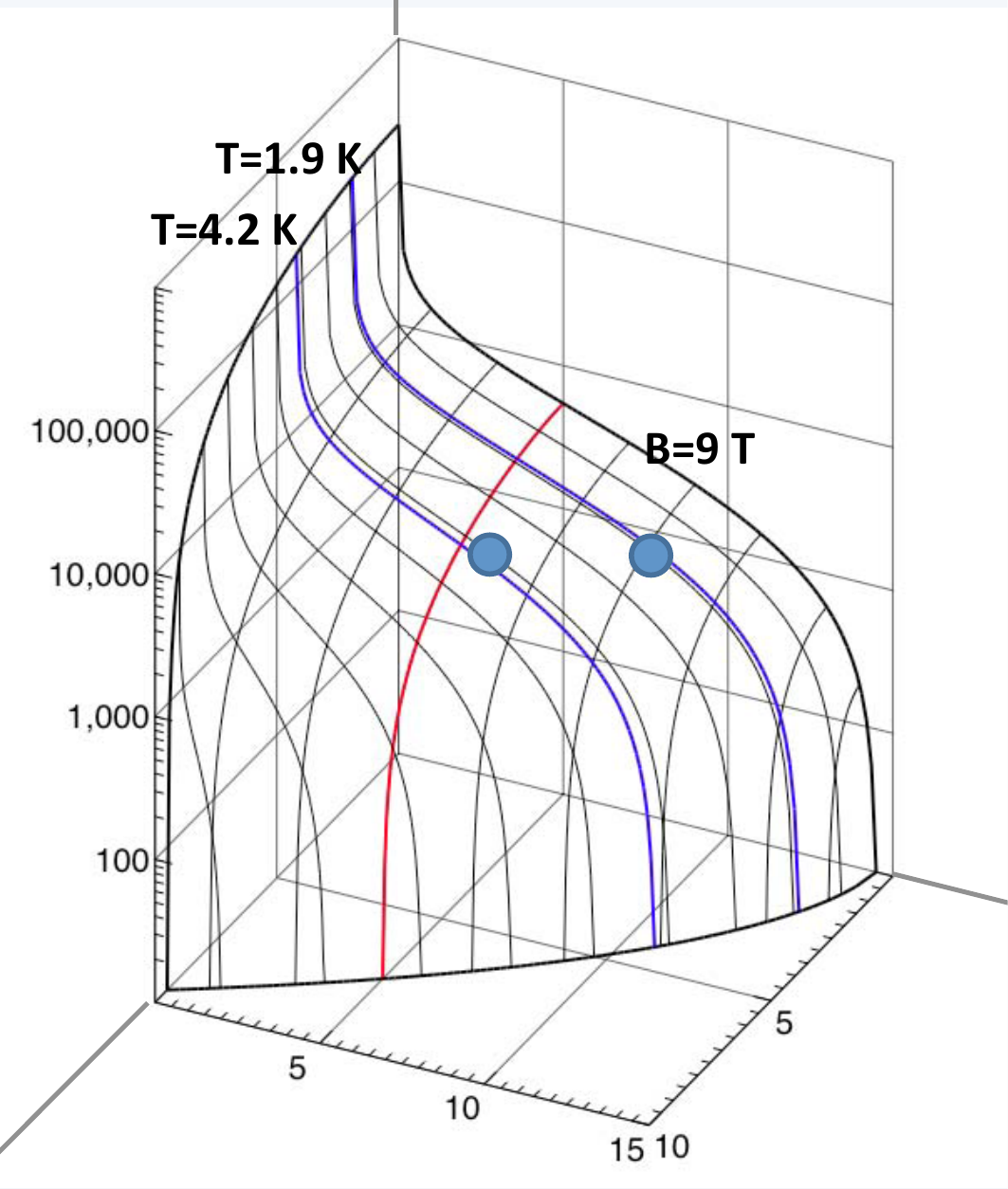} \hspace{1cm}
\includegraphics[height=5cm]{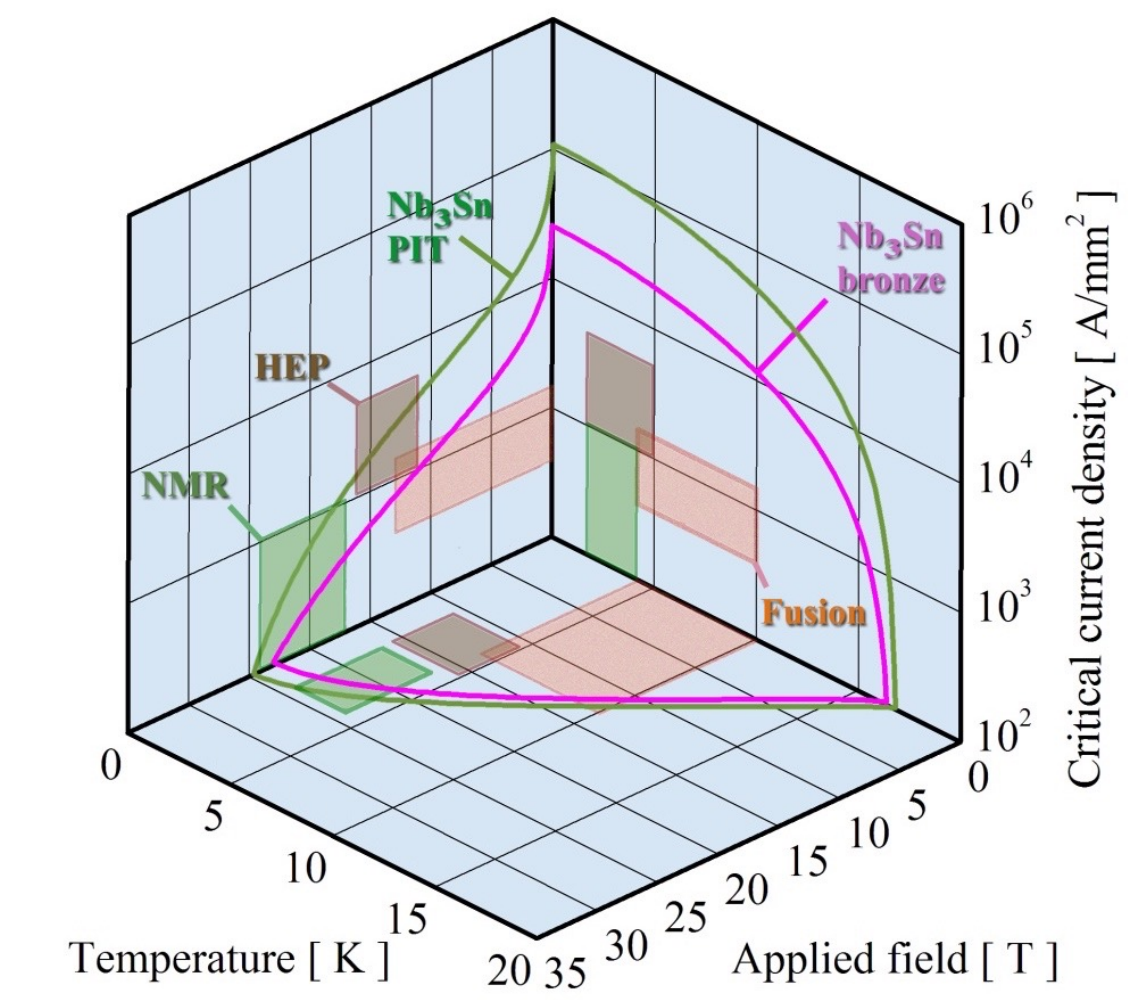}
\caption{Critical current surface in the Temperature--Flux density--current density parameter space for a Nb-Ti (left) and a Nb$_{3}$Sn (right) conductor }
\label{fig:critcur}
\end{center}
\end{figure}

We will now compare the two LTS conductors (Nb-Ti and Nb$_{3}$Sn) that are commonly available. In Figure~\ref{fig:critcur} typical critical surface plots for the two conductors can be found.

\begin{figure}[ht!]
\begin{center}
\includegraphics[height=6.5cm]{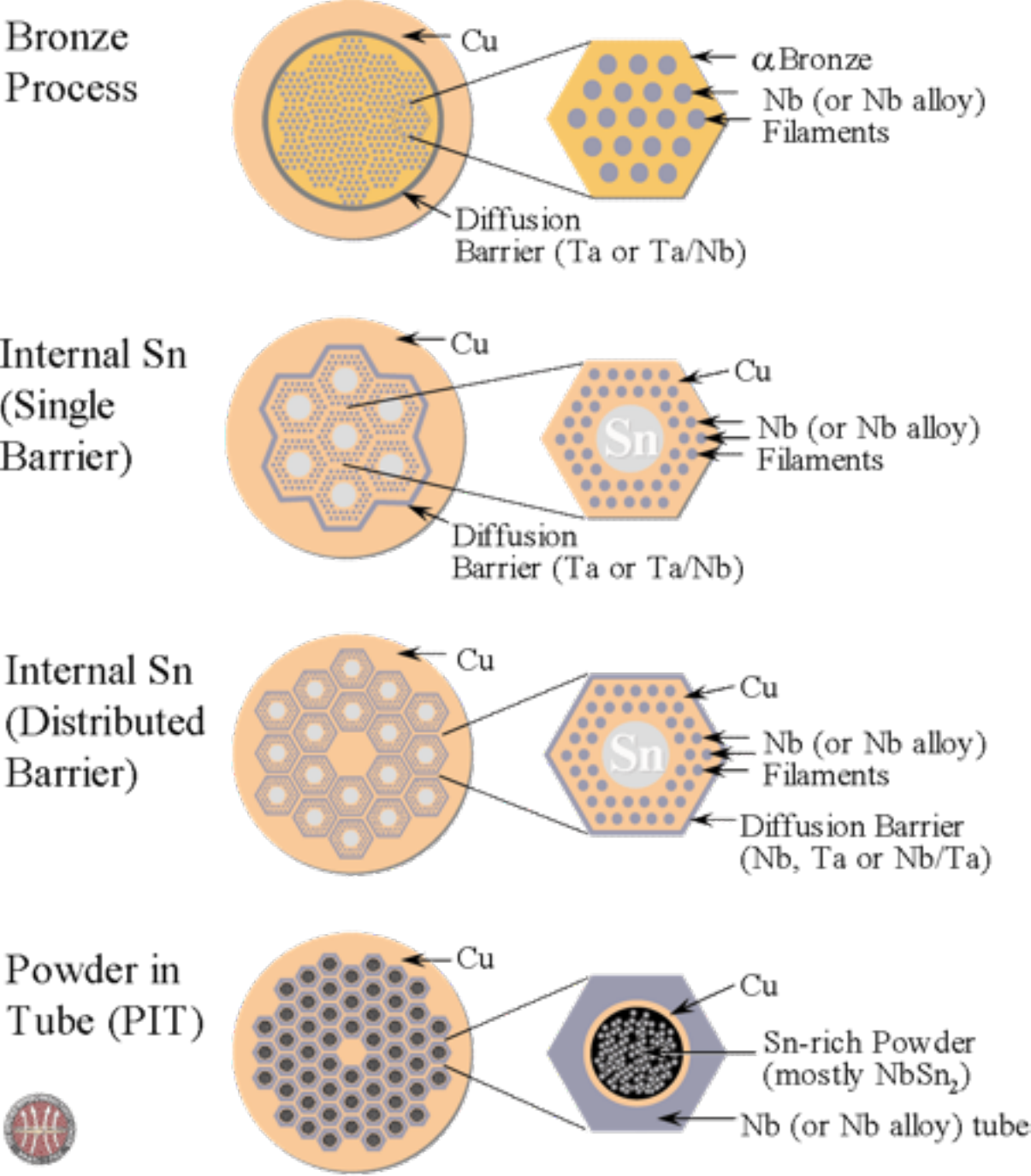}
\caption{Nb$_{3}$Sn wires types}
\label{fig:nb3sn}
\end{center}
\end{figure}

\begin{itemize}
\item{Nb-Ti} \\
This conductor is the workhorse for magnets between $B=4\UT$ and $B=10\UT$. It is produced in a well known industrial process. Thousands of accelerator magnets have been built using Nb-Ti conductors and $ B=10 \UT$ is the practical limit at  $T=1.9 \UK$. Niobium and titanium are combined into a ductile and mechanically very strong alloy. Bars of Nb-Ti are inserted into a hollow copper cylinder that is the basis of the wire (see Fig.~\ref{fig:nbti}). Starting from this billet, the wires are produced by successive extrusion and drawing steps and temperature treatments such that the desired wire geometry is obtained with its specific metallurgical structure. Filament diameter inside the wire is typically in the micron range. The critical temperature is $T_{c}=9.2\UK$ at $B=0\UT$ and the upper critical flux density is  $B_{c2}\approx14.5\UT$ at $T=0\UK$. The cost of a typical wire for accelerator magnets is approximately 100-150 US\$ per kg of wire. Critical currents of up to $J=2500 \UA/\UmmZ^{2}$ at $B=6 \UT$ and $T= 4.2 \UK$ or at $B=9 \UT$ and $T=1.9 \UK$ can be obtained.

\begin{figure}[ht!]
\begin{center}
\includegraphics[height=3.cm]{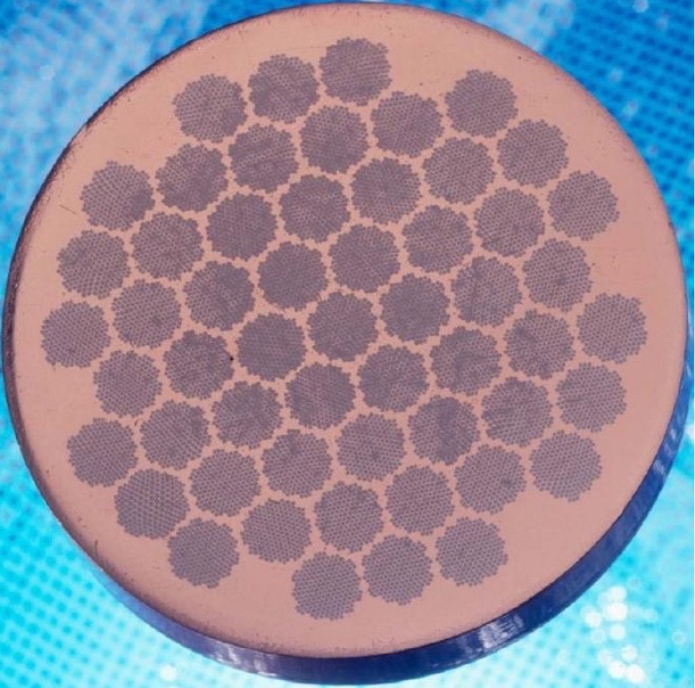} \hspace{1cm}
\includegraphics[height=3.cm]{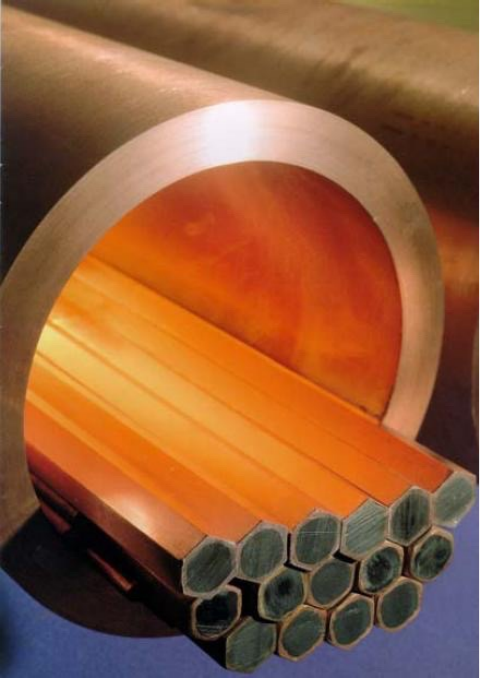}
\caption{Nb-Ti wires; left: Cross section of a wire, right: Nb-Ti billet}
\label{fig:nbti}
\end{center}
\end{figure}

\item{Nb$_{3}$Sn}\\
Nb$_{3}$Sn is the conductor available for magnets above $B=10\UT$ but limited to $B\simeq 16\UT$ for accelerator magnets. The industrial process to make the conductor is complex and has high costs. The conductor is brittle and strain sensitive and is therefore tricky to handle. 25+ short models for accelerator magnets and a few long prototypes have been built up to  now and $B=20 \UT$ is the ultimate limit at $T=1.9 \UK$, but above $B\simeq 16 \UT$ accelerator magnet coils will get too large and expensive.  Critical currents of up to $J=3000 \UA/\UmmZ^{2}$ at $B=12 \UT$ and $T= 4.2 \UK$ can be obtained. Niobium and tin form a A15 inter-metallic compound. This compound is brittle and strain sensitive. Bars or tubes of niobium with tin either inside or close-by are inserted into a~hollow copper cylinder, the billet, that is the basis of the wire. Different layout possibilities can be used for this, as can be seen in Fig.~\ref{fig:nb3sn}. Starting from the billet, the wires are produced by successive extrusion and drawing steps such that the desired wire geometry is obtained. The result is a wire with filaments of niobium with tin close-by in the form of powder, filaments or sheets. The wire has to be reacted at $T \approx650\UDC$ for around 100 hours to obtain the Nb$_{3}$Sn filaments inside the wire by a chemical reaction between the niobium and the tin. Filament diameter inside the wire are typically in the tens of micron range. The high current carrying Nb$_{3}$Sn is in the form of crystals with sizes of a few tens of $\UnmZ$. The critical temperature is $T_{c}=18\UK$ at $B=0\UT$ and zero strain and the upper critical flux density is  $B_{c2}\approx28\UT$ at $T=0\UK$ and zero strain. The cost of a typical wire for accelerator magnets is approximately 700-1500 US\$ per kg of wire.
\end{itemize}

% SUBSECTION
\subsection{High Temperature Superconductors}
Two types of HTS conductors are at the moment commercially available, they are bismuth strontium calcium copper oxide (BSCCO) and rare-earth barium copper oxide (ReBCO). The amounts produced are still modest and the production is still developing. They are not metallic but ceramic substances and they have a critical flux density above $100\UT$.

\begin{figure}[ht!]
\begin{center}
\includegraphics[height=4cm]{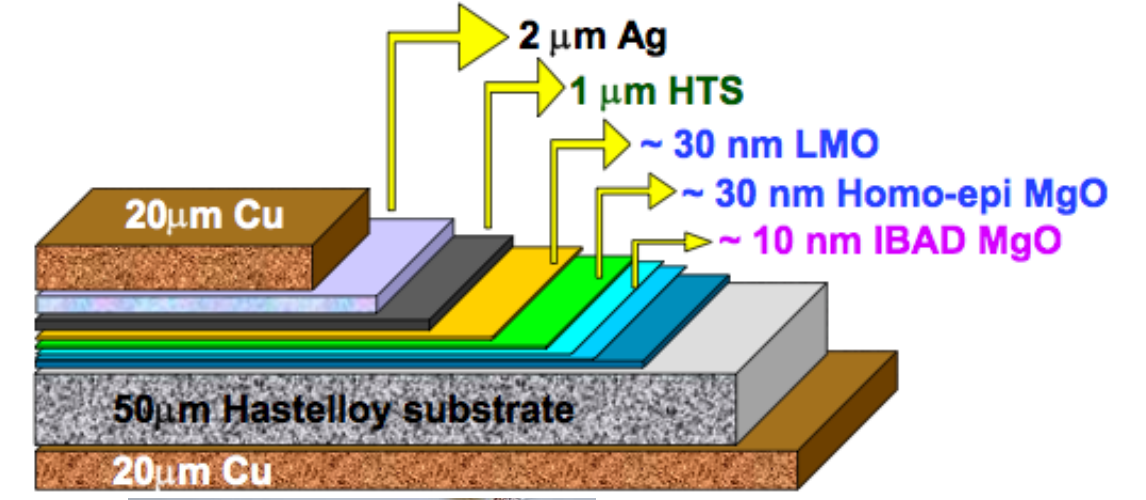}  \vspace{0.5cm}
\includegraphics[height=5cm]{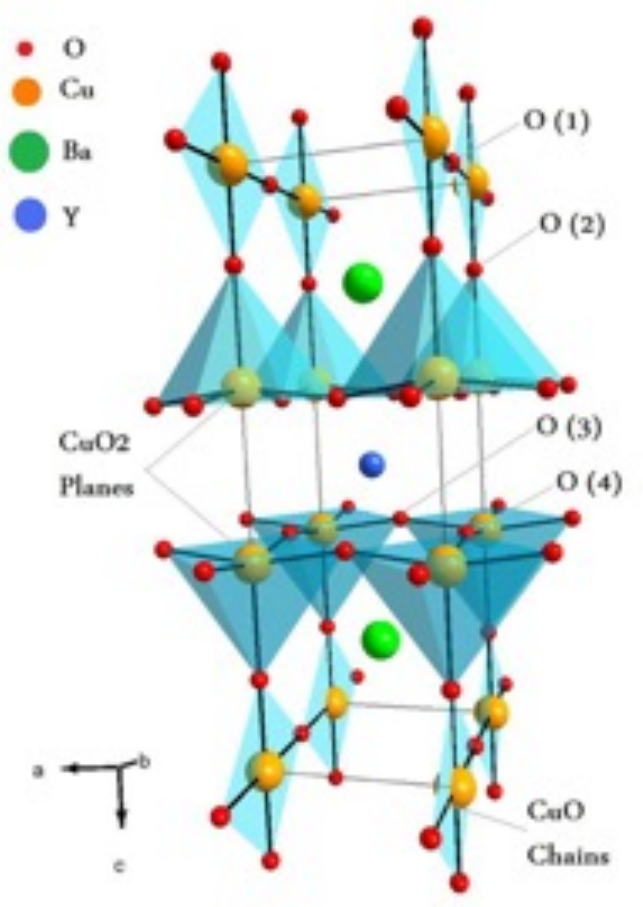}
\includegraphics[height=4cm]{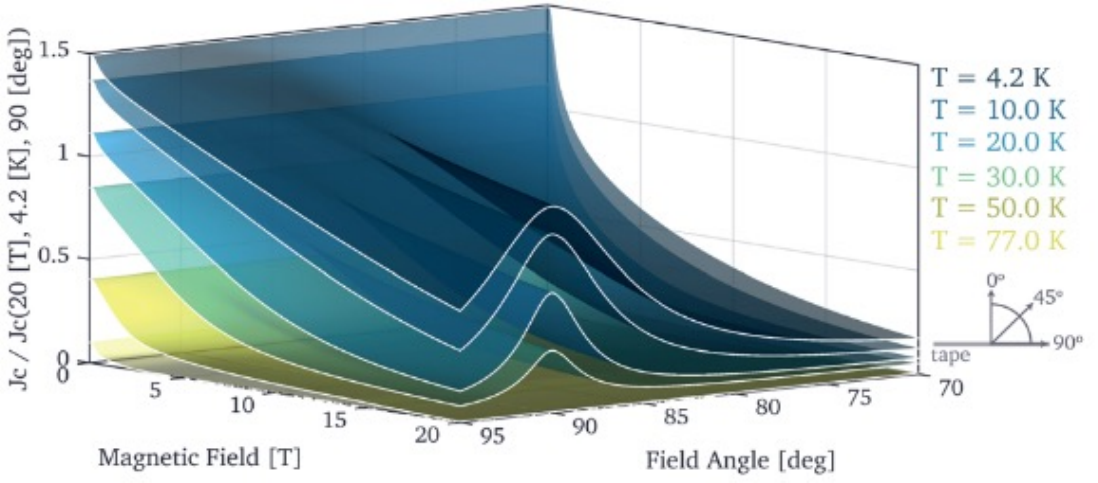}
\caption{ReBCO; left top: ReBCO tape, right top: Structure of ReBCO crystal, bottom: An example of field and field angle dependance of tape critical current}
\label{fig:ReBCO}
\end{center}
\end{figure}

\begin{itemize}
\item{BSCCO} bismuth strontium calcium copper oxide \\
BSCCO is a brittle substance and is, for accelerator magnet models, mostly used as a wind-and-react conductor where wires containing a pre-curser are wound into a coil and then reacted at $T\approx850 \UDC$. It is available in round wires where the BSCCO pre-curser filaments are situated inside a silver substrate (see: Fig.~\ref{fig:BSCCO}). The pre-curser does not yet contain oxygen and this is absorbed during the reaction for which the oxygen transparent silver substrate is needed. The~reaction is done in an atmosphere with $\approx1$ bar of oxygen gas and $\gtrsim20$ bar of an inert gas. It can reach an~overall critical current $J=1000\UA/\UmmZ^{2}$  at $B=20 \UT$ and $T=4.2 \UK$. The round wires can be cabled into a Rutherford cable (details on Rutherford cables in \Sref{sec:Rutherford}). BSCCO is a~difficult technology due to the reaction parameters and the strain sensitivity, but could be promising for high field magnets in $B\geq16\UT$ region. Prices are still much higher than for LTS conductors but this conductor is still being developed.

\begin{figure}[ht!]
\begin{center}
\includegraphics[height=5cm]{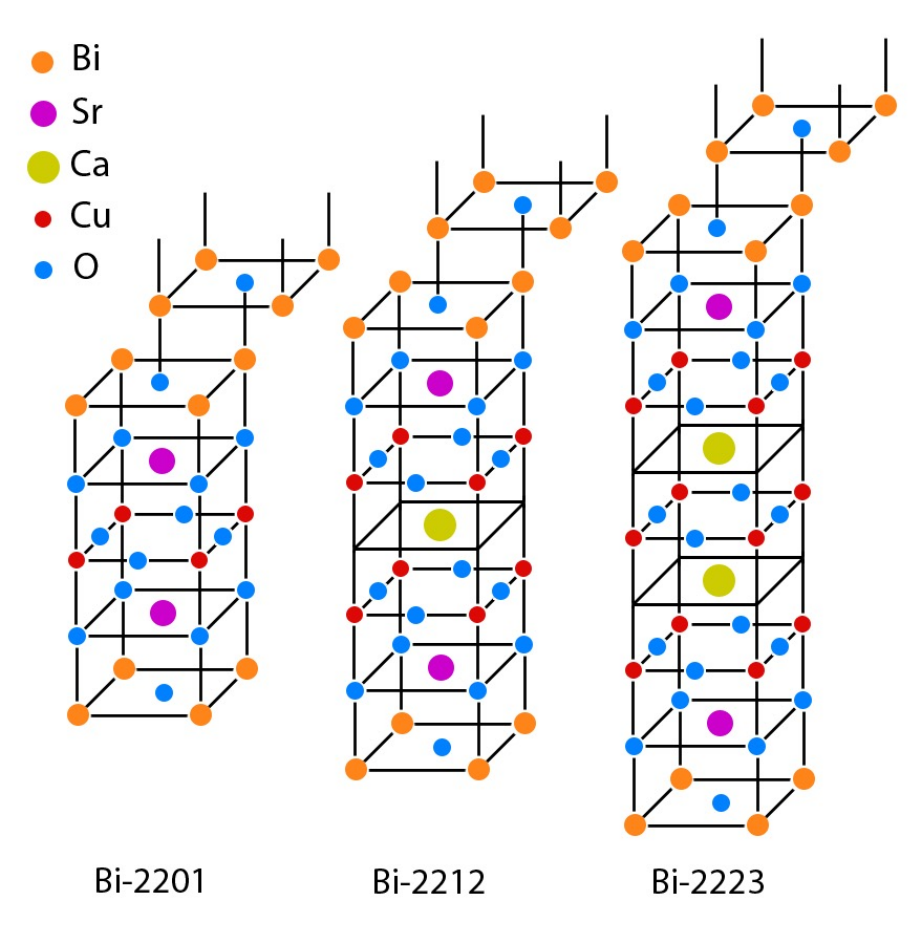} \hspace{1cm}
\includegraphics[height=4cm]{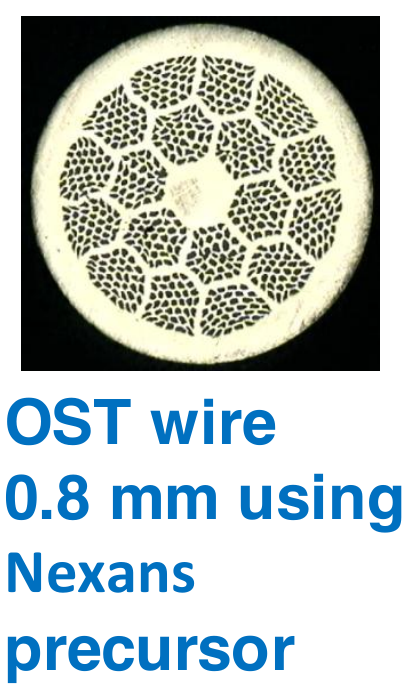}
\caption{BSCCO; left: structure of BSCCO crystals, right: BSCCO-2212 wire}
\label{fig:BSCCO}
\end{center}
\end{figure}

\item{ReBCO} Rare-earth barium copper oxide \\
ReBCO conductor comes in the form of a tape. The basis of the tape is a stainless steel or Hasteloy substrate with a thickness varying between 30 and 100 $\UumZ$ and a width between 4 and 12 $\UmmZ$. On one side of the substrate several layers of crystaline substances are deposited to provide a specific texture. On this the ReBCO layer is then deposited. The ReBCO layer can have a thickness between 1 and 3 $\UumZ$. On top of the ReBCO a thin silver layer is put as protection. The tape is enrobed with a layer of copper with a thickness of a few tens of $\UumZ$ for mechanical and chemical protection and to serve as a current shunt.  The tape is mechanically very robust and can be subjected to axial and transversal stress of a few hundred M\UPaZ. It can reach an overall critical current density $J=900 \UA/\UmmZ^{2}$  at $B=20 \UT$ and $T=4.2 \UK$, but the critical current depends on the angle between the face of the tape and the magnetic field (see Fig.~\ref{fig:ReBCO}). The tapes have only limited cabling possibilities. ReBCO is a difficult technology, due to the cabling layout, but could be promising for high field magnets in $B\geq16\UT$ region. Prices are still much higher than for LTS conductors but this conductor is still being developed and one can already see a steady decrease in the last few years.
\end{itemize}

% SUBSECTION
\subsection{Superconducting cables for magnets}
\label{sec:Rutherford}
\begin{figure}[tb!]
\begin{center}
\includegraphics[height=5.cm]{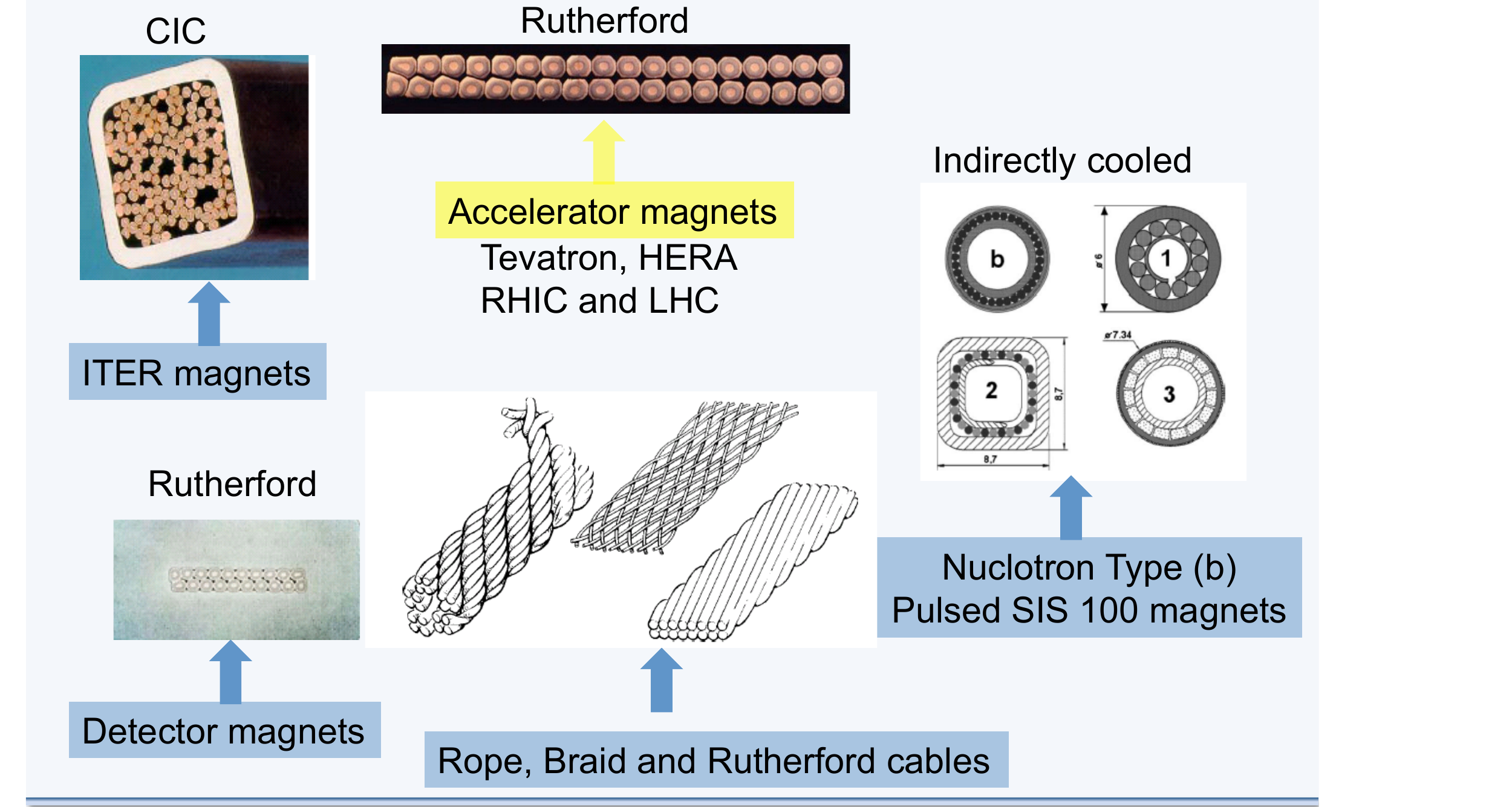} \hspace{0.5cm}
\includegraphics[height=6cm]{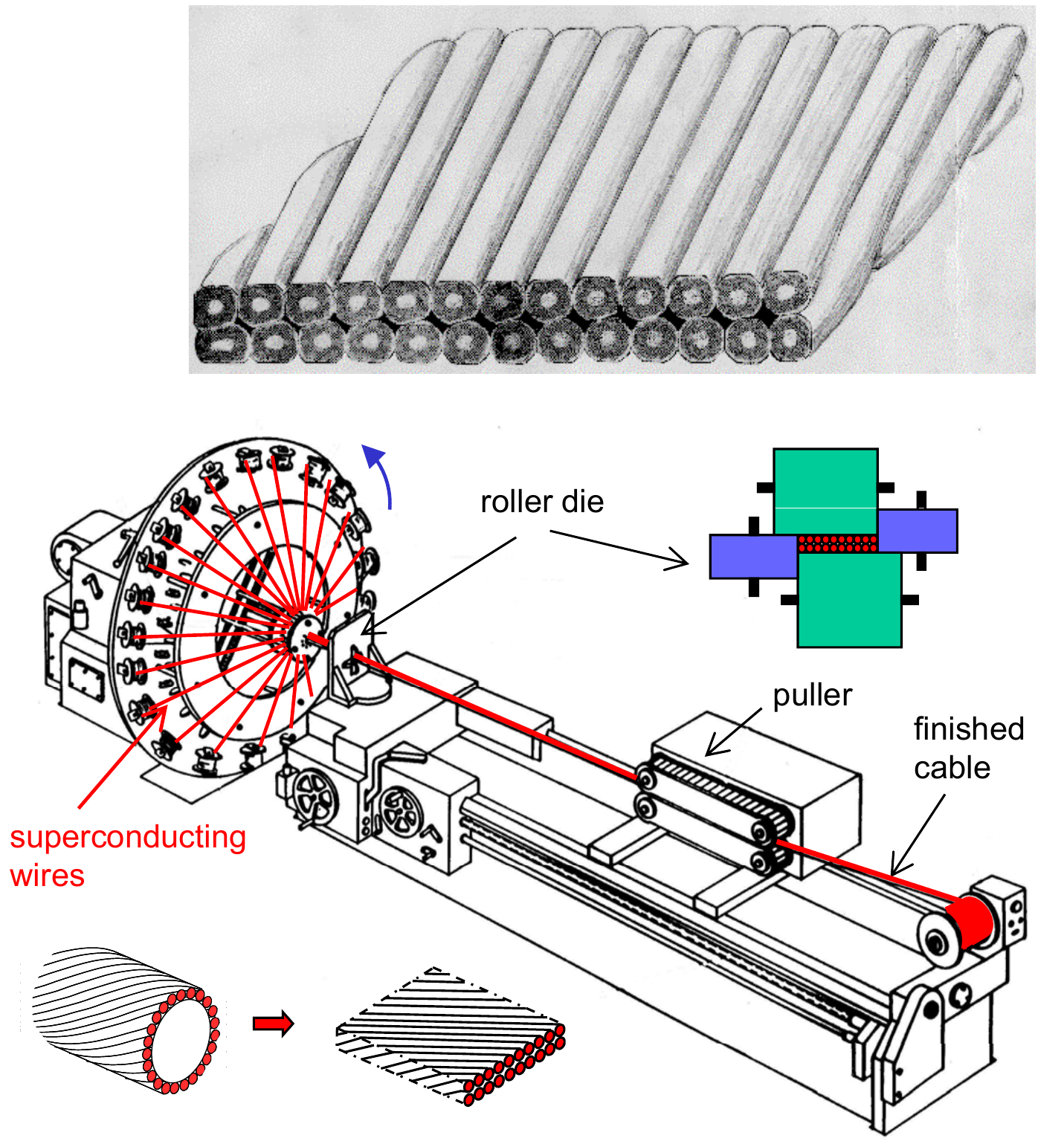}
\caption{left: LTS cables, right: Rutherford cabling machine}
\label{fig:cables}
\end{center}
\end{figure}

Accelerators with superconducting magnets have typical energy ramp-up times between 100$\Us$ and 1000$\Us$. The inductive voltage over a magnet during a current ramp is 
$V=-L\frac{dI}{dt} $, while the inductance of the~magnet is related to the number of turns in the coils by $L\sim N^{2}$. For practical reasons, magnets are designed with an insulation system that is rated for a maximum coil to ground voltage of 1000$\UV$. One thus has to take care that the combination of the inductance and the current ramp-rate keeps the inductive voltage below this value. This can only be done by limiting the number of turns in the coil. As $B\sim N\cdot I$ we need to use an appropriately high current conductor. In table ~\ref{tab:acccurr} we can find the currents used for the main dipoles of three accelerators.  For future machines with $10 \UT<B<15\UT$ the currents will have to be $10\UkA<I<15\UkA$. For stability reasons the wires of LTS conductors have a diameter $0.6\Umm<d<1\Umm$. With a copper-to-non-copper material volume ratio in the wire of around 1 and a critical current density in the superconductor $J_{c} \approx1000 \UA / \Umm^{2}$ a 1 $\Umm$ diameter wire can carry $\approx 400\UA$. This means that we need a cable with 30 wires to reach a current of 12$\UkA$. To build compact accelerator magnets with LTS conductor the preferred cable type is a Rutherford cable (see Fig.~\ref{fig:cables} ).

\begin{table}[h]
\begin{center}
\caption{Current of the dipoles in superconducting accelerators.}
\label{tab:acccurr}
\begin{tabular}{|p{3cm}|c|c|}
\hline\hline
\textbf{Machine}    & \textbf{B (T)}  & \textbf{I(A)}\\
\hline
Tevatron    &     4.4    &  $\approx$4000 \\
Hera          &      5      &  $\approx$6000 \\
LHC          &      8.3    & $\approx$12000\\
\hline\hline
\end{tabular}
\end{center}
\end{table}

% SECTION
\section{Practical magnet design and manufacturing}
As a first step in the design of a magnet we want to do a first order estimation of the cross section and for this we should make an estimation of the coil size.  For this we can use scaling laws. These scaling laws can be found in Ref. \cite{bib:Todesco1}, and indirectly in the books of Wilson \cite{bib:Wilson}, Mess \cite{bib:Mess} and Russenschuck \cite{bib:ROXIE}. As input parameters we need to know the desired field value, the current density of the conductor and the~aperture.

% SUBSECTION
\subsection{Dimensioning of the coils, scaling laws}

\begin{figure}[ht!]
\begin{center}
\includegraphics[height=2.5 cm]{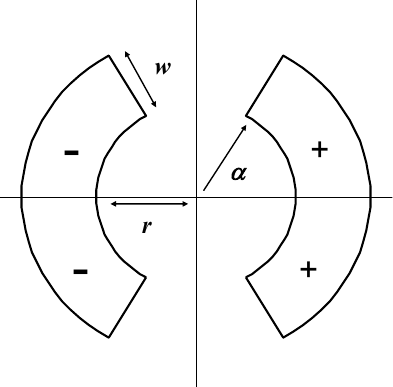} \hspace{0.5cm}
\includegraphics[height=2.5 cm]{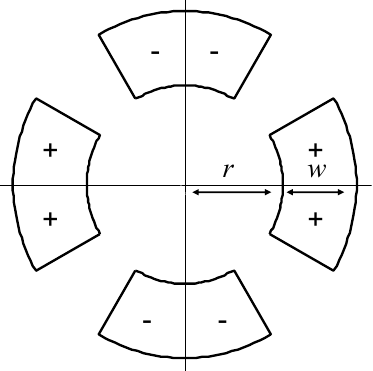}
\caption{ left: Dipole sector coil; right: Quadrupole sector coil}
\label{fig:sectorcoils}
\end{center}
\end{figure}

\subsubsection{Dipole coil width}
From Ampere's law we can derive the field resulting from the current in a line conductor and integrate this over the surface of a $60^{\circ}$ dipole sector coil (see Fig.~\ref{fig:sectorcoils})

\begin{equation}
B_{1} = -4  \frac{j\mu_{0} }{2\pi}\int^{\pi/3}_{0} \int^{r+w}_{r}\frac{\cos(\theta) }{\rho}\rho d\rho d\theta = - \frac{\sqrt{3}\mu}{\pi} jw
\end{equation}

with: $r$ coil inner radius; $w$ coil width; $\rho$ radial coordinate; $j$ current density. 
We can see that:

\begin{itemize}
\item{the field is proportional to the current density $j$}
\item{the field is proportional to the coil width $w$}
\item{the field is independent on the aperture diameter.}
\end{itemize}
With this we can sketch out a cross section in the case of a sector coil or $\cos\Theta$ coil. The coil width will be roughly the same for the case of a block coil.

\subsubsection{Quadrupole coil width}
In the same way we can derive from Ampere's law the field resulting from the current in a line conductor and integrate this over the surface of a $30^{\circ}$ quadrupole sector coil (see Fig.~\ref{fig:sectorcoils})

\begin{equation}
G = -8  \frac{j\mu_{0} }{2\pi}\int^{\pi/6}_{0} \int^{r+w}_{r}\frac{\cos(\theta) }{\rho}\rho d\rho d\theta = - \frac{\sqrt{3}\mu}{\pi} j \ln \left(1+\frac{w}{r} \right).
\end{equation}

We can see that:
\begin{itemize}
\item{the field gradient is proportional to the current density $j$}
\item{the field gradient depends on $w/r$.}
\end{itemize}

This will allow us to draw out a first cross section of the quadrupole.

% SUBSECTION
\subsection{Electromagnetic forces and stress}

\begin{figure}[ht!]
\begin{center}
\includegraphics[height=6 cm]{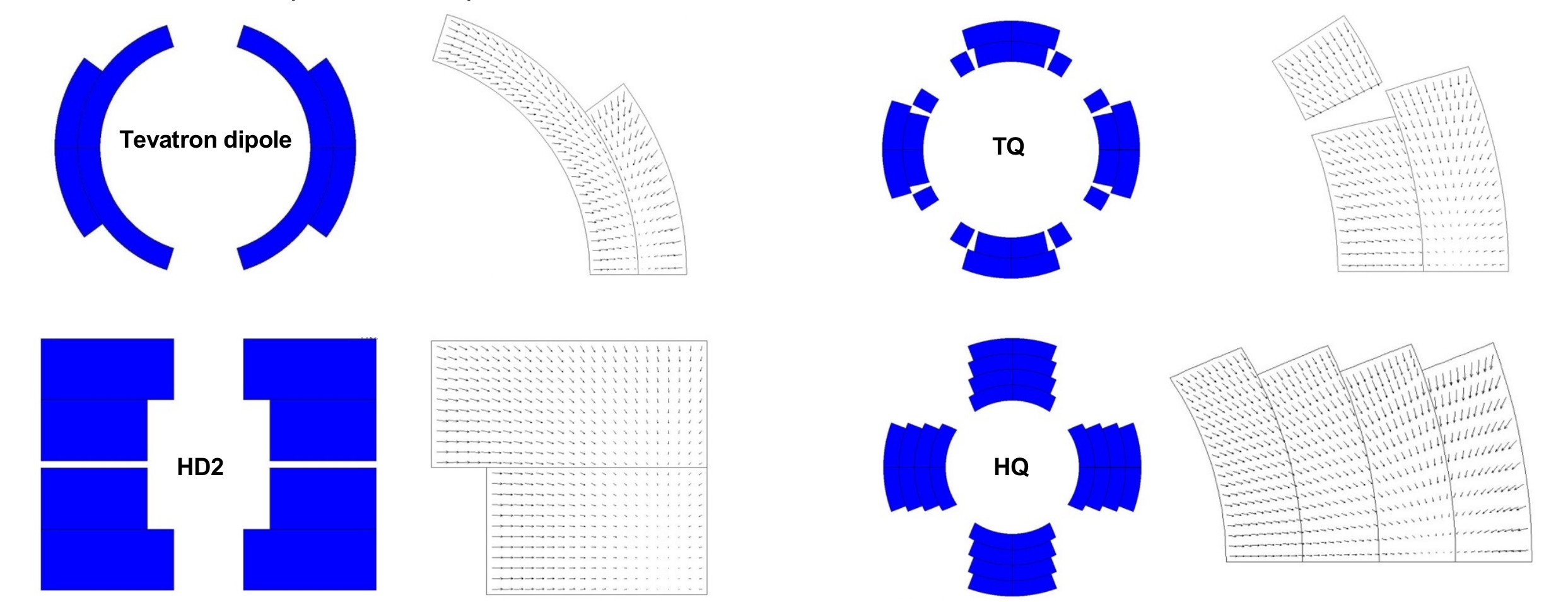} 
\caption{Coil shapes and electromagnetic forces. Left top: Dipole $\cos\Theta$ coil (Tevatron); left bottom: Dipole block coil (HD2); right top: Quadrupole $\cos\Theta$ coil; (T) right bottom: Quadrupole multi layer $\cos\Theta$ coil (HQ).}
\label{fig:coilforces}
\end{center}
\end{figure}

In Figure~\ref{fig:coilforces} we can see for some coil types the resulting electromagnetic forces. For a dipole magnet in case of a $\cos\Theta$ coil the forces have an outward component and a component towards the mid-plane. As a result there is an important stress concentration on the coil mid-plane. For a dipole block coil, the~coil forces predominantly point outwards. For quadrupoles, essentially only $\cos\Theta$ coils are used and also here the forces have an outward component and a component towards the mid-plane resulting in an~important stress concentration on the coil mid-plane.
Using scaling laws we can make an estimate for the maximum stress caused by the electromagnetic forces in the coil.

\subsubsection{Maximum stress in a dipole coil}
Using the book of Wilson \cite{bib:Wilson} and the paper of Fessia et al \cite{bib:Fessia1}, we can derive that for a $60^{\circ}$ dipole sector coil the~maximum stress is:

\begin{equation}
\sigma \approx j^{2} \frac{\mu_{0} \sqrt{3}} {6 \pi} \max \hspace{0.02cm} _{\rho\epsilon\left[r,r+w\right]} \left[  2\rho^{2} + \frac{r^{3}}{\rho} - 3\rho\left(r+w\right)  \right]
\end{equation}
with: $\sigma$ the maximum stress; $r$ coil inner radius; $w$ coil width; $\rho$ radial coordinate; $j$ current density. 
For an $8\UT$ dipole magnet this is typically 40 M$\UPaZ$ and for a $13\UT$ magnet 130 M$\UPaZ$.

\subsubsection{Maximum stress in a quadrupole coil}
Using again the book of Wilson \cite{bib:Wilson} and the paper of Fessia et al \cite{bib:Fessia2}, we can derive that for a $30^{\circ}$ quadrupole sector coil the maximum stress is:

\begin{equation}
\sigma \approx j^{2} \frac{\mu_{0} \sqrt{3}} {16 \pi} \max \hspace{0.02cm} _{\rho\epsilon\left[r,r+w\right]} \left[  2\rho^{2} + \frac{r^{4}}{\rho^{2}} + 4\rho^{2}\ln\left(\frac{r+w}{\rho}\right)  \right].
\end{equation}

\newpage
\section{Conductor stability and dynamic field quality}
\begin{figure}[ht!]
\begin{center}
\includegraphics[height=1.5 cm]{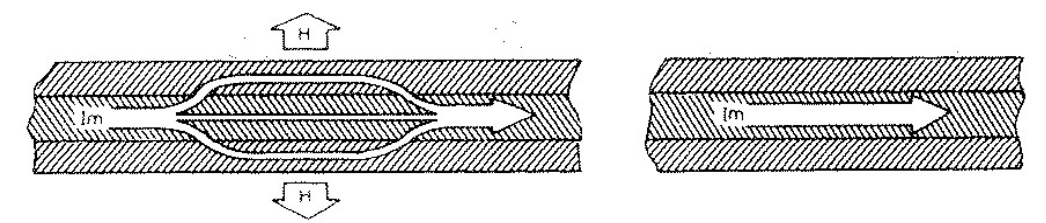} \hspace{1.5 cm}
\includegraphics[height=3 cm]{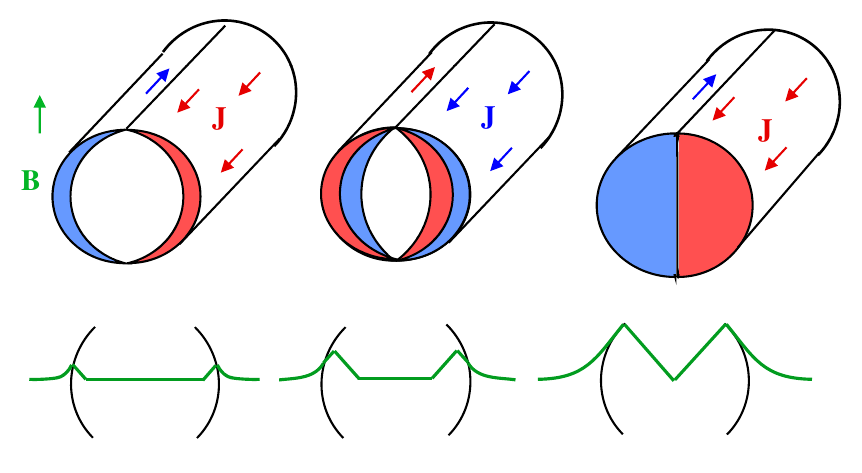} 
\caption{left: Cryostabilazation of a superconductor by surrounding it with Cu; right: Filament magnetisation with shielding currents}
\label{fig:stabil}
\end{center}
\end{figure}

\begin{figure}[ht!]
\begin{center}
\includegraphics[height=2.5 cm]{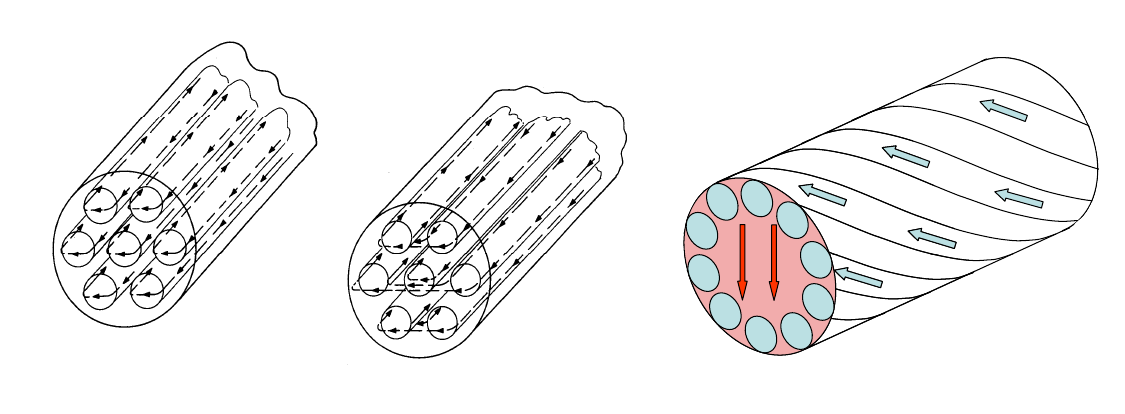}  
\caption{Twisting wires with filaments to reduce magnetic effects}
\label{fig:twist}
\end{center}
\end{figure}

Pure LTS superconductors are not stable as superconductors are poor normal conductors due to the low number of "free" conduction electrons in the lattice. To remedy this problem the conductors need to be cryogenically stabilized by surrounding them with a good normal conductor like copper (or aluminium). In Figure~\ref{fig:stabil} we can find an illustration of the current shunted through the copper surrounding a superconducting filament. During the ramping up of the current, shielding currents along the filaments will occur that generate unwanted fields (see Fig.~\ref{fig:stabil}) . These can be reduced by making the filaments very thin (a~few microns for Nb-Ti). To further reduce the magnetic effects of the shielding currents in the filaments, the wires are twisted  (see Fig.~\ref{fig:twist}). 
The final result is a wire that consists of many (from a~few hundred to a few thousand) thin filaments of superconductor in a copper matrix that is twisted.

\section{Quench, a thermal run-away effect}
\begin{figure}[ht!]
\begin{center}
\includegraphics[height=3.5 cm]{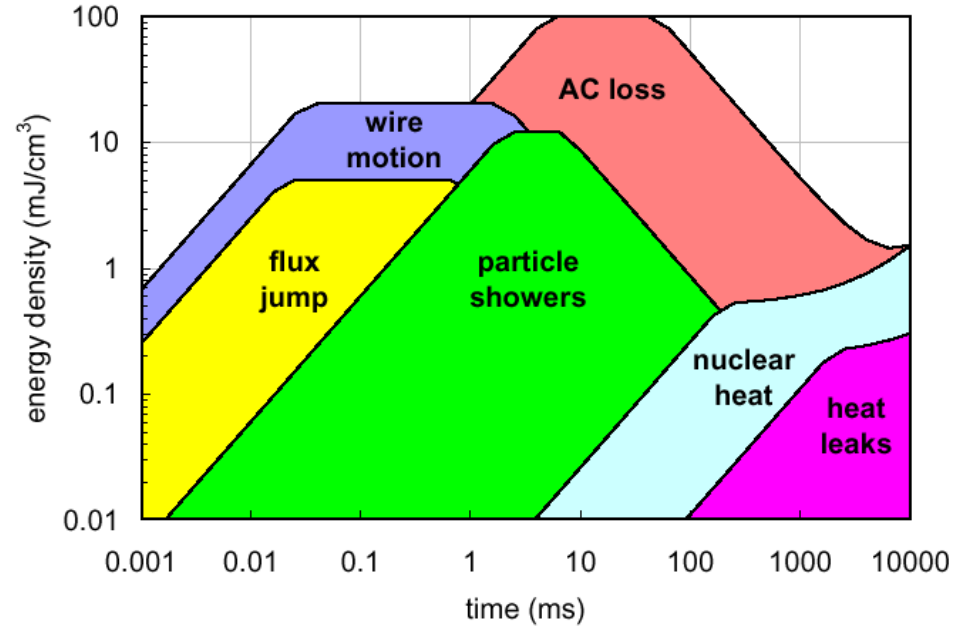} \hspace{0.2 cm}
\includegraphics[height=3.5 cm]{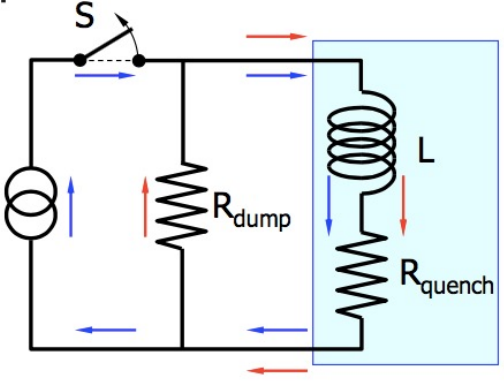} \hspace{0.2 cm}
\includegraphics[height=4 cm]{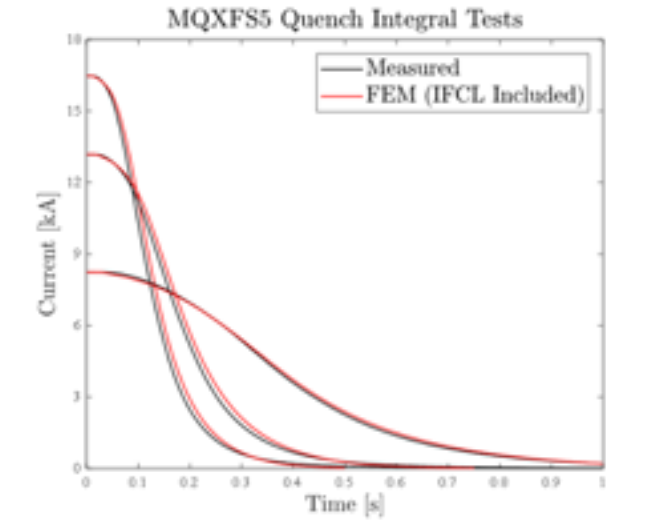}  
\caption{left: Energy spectrum of perturbing effects that can lead to a quench; middle: typical powering circuit of a superconducting magnet; right: Current decay after a quench for an MQXF quadrupole short magnet}
\label{fig:quench}
\end{center}
\end{figure}

Due to perturbations or when the current limit is locally reached, the conductor can locally get over the~critical current $ T > T_{c} (J,B)$. If the resistive heat that is generated exceeds the cooling then a~thermal runaway will occur called  "quench". In Figure~\ref{fig:quench} one can see a spectrum of energy versus duration of various perturbation sources. Quenches can be caused by slow low heating effects and by very short spikes. With stored energies in the >$\UMJZ$ range this can potentially overheat the coil locally. Temperatures of T $\approx3000\UK$ are possible. Such events can destroy the coil and even the whole magnet. Superconducting magnets thus need a quench protection system that can be based on several protection measures in parallel. A typical electrical circuit for a superconducting magnet can be seen in Fig.~\ref{fig:quench}.\\
A standard sequence of actions in case of a quench is:

\begin{itemize}
\item Detect the quench. Measure the voltage over the magnet or an individual coil. A fully superconducting coil has $R=0 \rightarrow V=0$. When part of a coil is in the normal state $V>0$. Typically a~threshold of $100\UmV$ is used to trigger the protection measures.
\item Switch off the power converter.
\item Dump the energy of the circuit into a dump resistor by opening a switch.
\item Heat up the entire coil with quench heaters. This can be done by discharging a capacitor over a~thin metal strip that is put against the coil.
\end{itemize}

In Figure ~\ref{fig:quench}\ we can see an example of the current decay of a superconducting quadrupole.

\subsection{Quench hot spot}
When a quench starts locally in a coil, that spot generates heat due to the normal resistance. We can estimate the temperature that will be reached in this spot (the so called "hot spot") with a first order calculation. For this we assume that in the time scales involved ($t < 1\Us$) the coil cannot evacuate any heat (adiabatic approach).
The heat balance on a unit volume of coil is:

\begin{equation}
J^{2}(t)\rho(T)dt = \gamma C(T)dT
\end{equation}
with: $t$ time, $T$ temperature, $J(t)$ current density as function of time, $\rho(T)$ resistivity of the non-superconducting part of the conductor, $\gamma$ density, $C(T)$ heat capacity.
We need to know or assume the current decay function $J(t)$. We can now rearrange the expression:

\begin{equation}
J^{2}(t)dt = \frac{\gamma C(T)}{\rho(T)}dT   \mathrm{\hspace{0.5cm} Integrate: \hspace{0.5cm}} \int_{0}^{\infty} J^{2}(t)dt = \int_{T_{0}}^{T_{max}} \frac{\gamma C(T)}{\rho(T)}dT.
\end{equation}
The resulting equation then has to be solved to get $T_{max}$. With this approach one obtains a conservative estimate of the maximum temperature in the coil at the end of a quench.

% SECTION
\section{Practical coil geometries for superconducting magnets}
There are three types of coil geometries in use or foreseen for most superconducting accelerators: $\cos \theta$ coils, Block coils and Canted $\cos \theta$ coils. 

\subsection{$\cos \theta$ coils}
As was shown in Section 1.5, a perfect $\cos \theta$ layout of a coil will result in a perfect dipolar field with no other multipolar components, similarly a layout with $\cos \mathrm{n}  \theta$ generates a perfect n-polar field. In practice such a layout has to be made with a physical cable which will thus result in small multipolar components in the field. Nevertheless, for thin coils (W$_{coil} \ll \varnothing _{aperture}$) the $\cos \theta$ layout allows for a very good field quality ($b_{n} \lesssim10^{-4}$) while yielding the best possible efficiency in the quantity of conductor used. To reduce the field perturbations caused by non-ideally fitting cables, on a small inner coil diameter a~fitting wedge-shaped "keystoned" cable is not always possible to make, wedges are inserted in the coil. The position and size of such wedges are optimised with dedicated optimisation algorithms in the field modelling software (e.g. with the ROXIE \cite{bib:ROXIE} software package). The coil ends can be made relatively short. For the design of the coil ends there is a large experience although it is not easy due to the special geometry that depends on the flexibility parameters of the cable. $\cos \theta$ coils feature stress accumulation on the mid plane of the magnet (see Section 4.2) which for high field magnets made with brittle and stress sensitive conductors like Nb$_{3}$Sn can become a limiting factor. 

\subsection{Block coils}
Coils with a block geometry will not automatically yield a perfect dipolar field. The field homogeneity is though getting better in the case of wide coils (W$_{coil} \geq \varnothing _{aperture}$) and can be optimised by including wedges in the coil. The quantity of conductor that is used in a block coil can be up to $10\%$ higher than for an equivalent $\cos \theta$ coil. The geometry of the straight parts is very easy to design and fabricate. Coil ends are so-called "flared-ends" that use the flexibility of the cable in the hard-way bend direction and are thus somewhat longer than for $\cos \theta$ coils. Flared-ends are relatively easy to make but there is still a~limited experience in finding the most optimum mechanical parameters. Stress buildup is at the outside edge of the coil where the field is relatively low, giving a clear advantage for using brittle and stress sensitive conductors like Nb$_{3}$Sn. 

\subsection{Canted $\cos \theta$ (CCT) coils}
\begin{figure}[tb!]
\begin{center}
\includegraphics[height=3.4 cm]{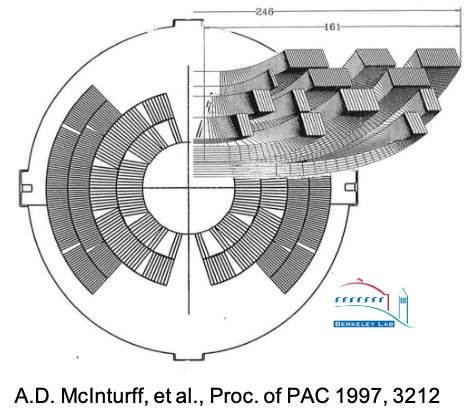} \hspace{0.5 cm}
\includegraphics[height=3.4 cm]{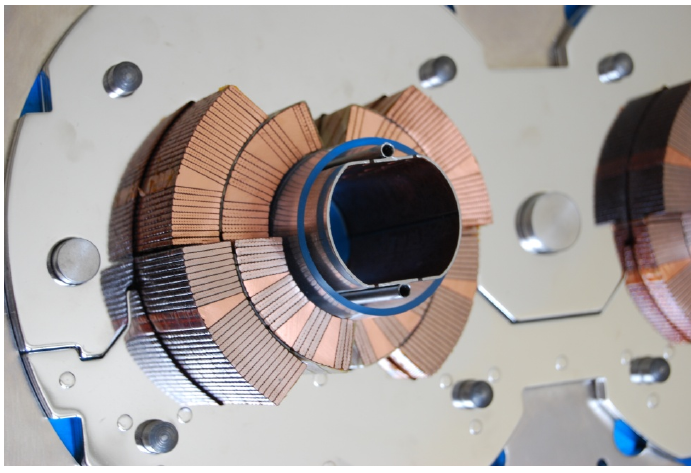} \hspace{0.5 cm}
\includegraphics[height=3.4 cm]{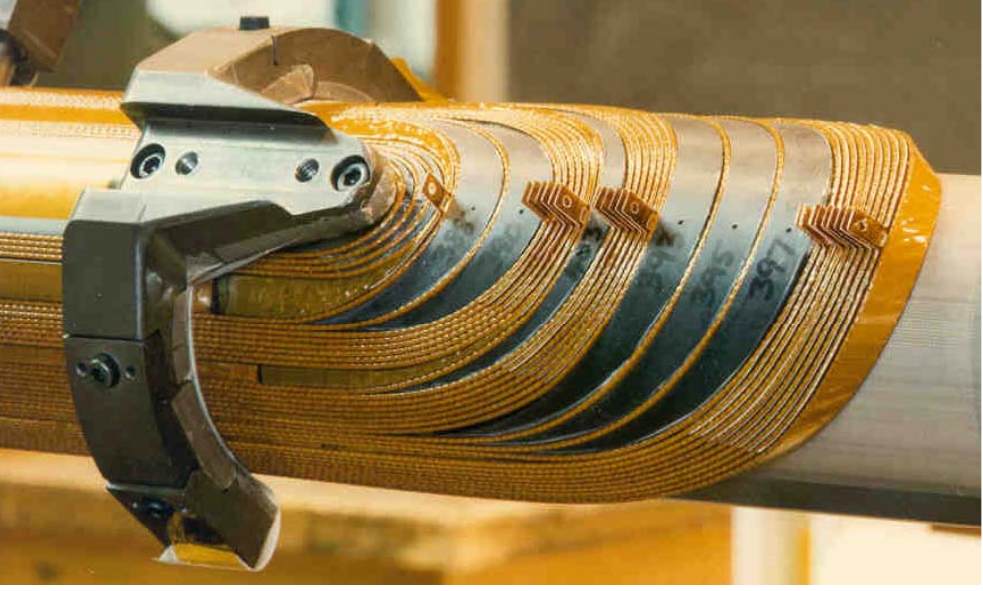} \hspace{0.5 cm}
\includegraphics[height=3.5 cm]{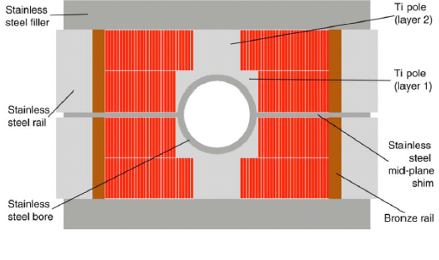} \hspace{0.5 cm}
\includegraphics[height=3.5 cm]{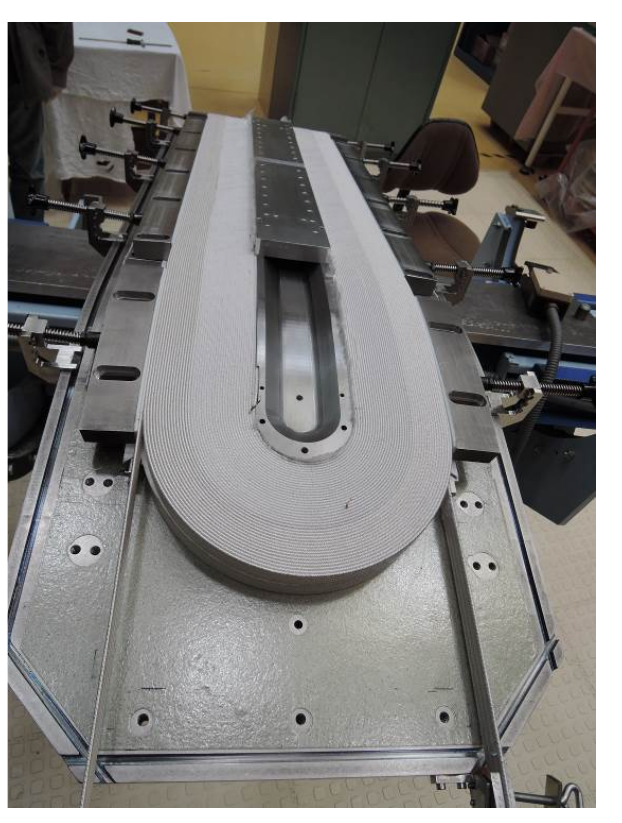} \hspace{0.5 cm}
\includegraphics[height=3.5 cm]{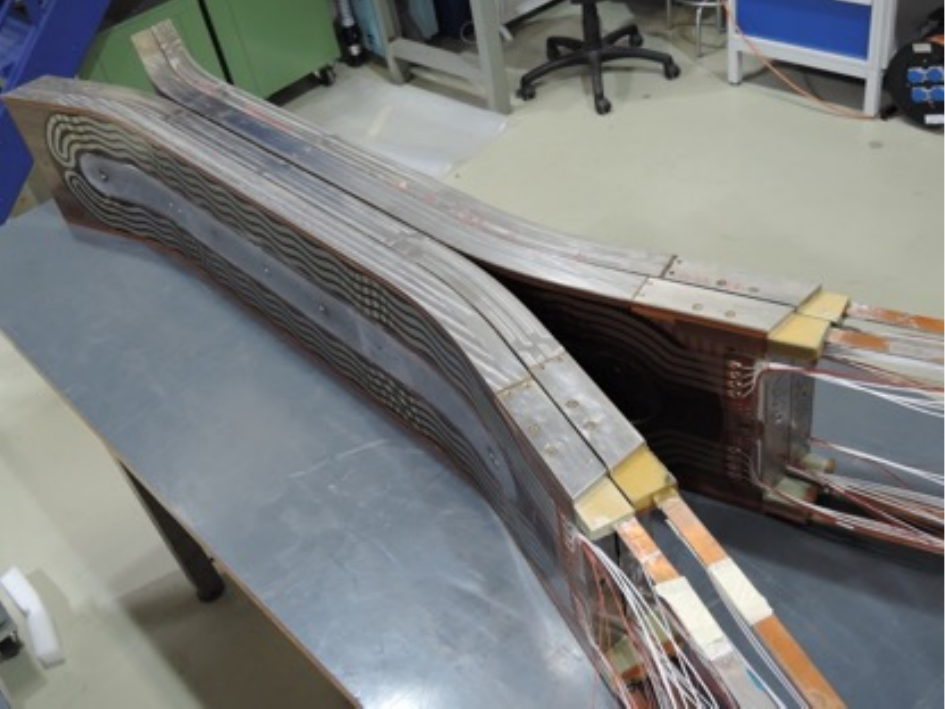} \hspace{0.5 cm}
\includegraphics[height=3.5 cm]{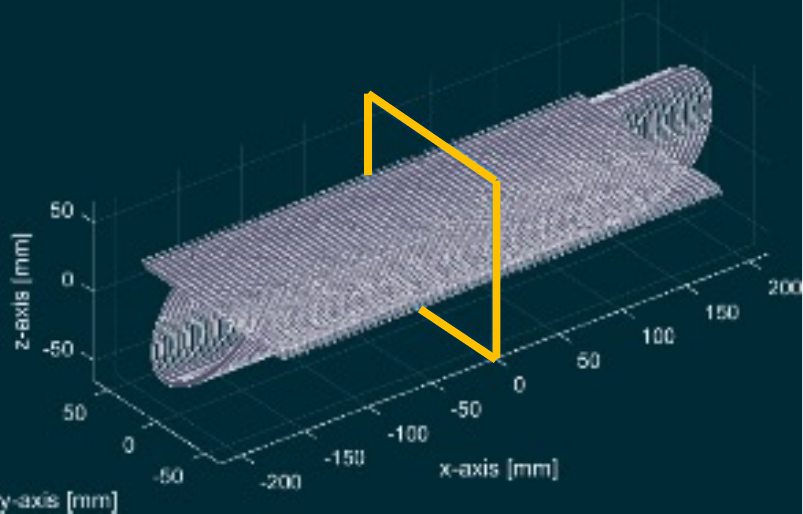} \hspace{0.5 cm}
\includegraphics[height=3.5 cm]{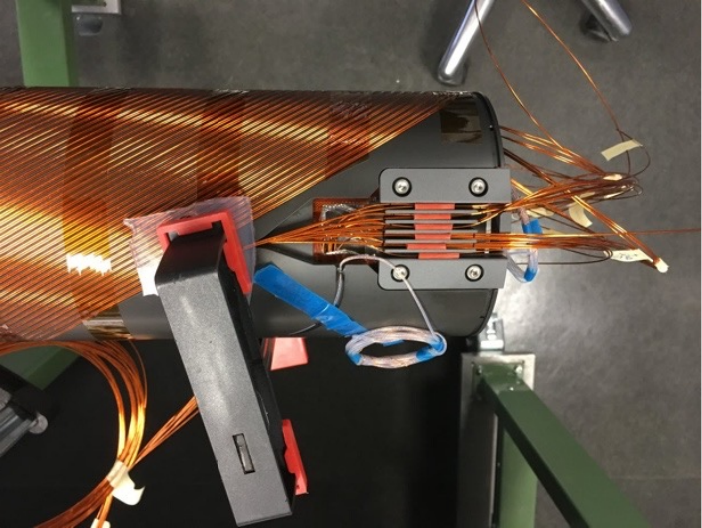} \hspace{0.5 cm}
\includegraphics[height=3.5 cm]{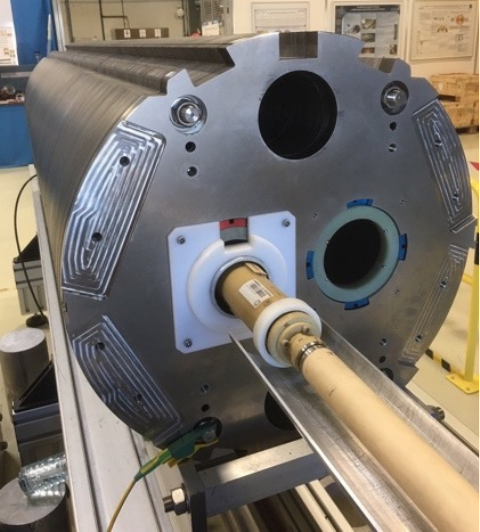} 
\caption{Top row: $\cos \theta$ coils; left: D20 (LBNL), middle: LHC main dipole, right: Coil end. 
Middle row: Block coils; left:  HD2 (LBNL) coil pack, middle: Fresca2 coil on winding table, right: finished Fresca2 coils. 
Bottom row: CCT; left: CCT dipole coils, middle: CCT coil connections, right: HL-LHC MCBRD at CERN.}
\label{fig:coilgeom}
\end{center}
\end{figure}

Canted $\cos \theta$ magnets are based on an idea that is more that 40 years old and has only recently been picked up again to be employed in high energy accelerators. The magnets consists of two concentric solenoids that are canted in opposite directions and powered in opposite polarities. The result is that the~axial field component of the two coils is cancelling and the transverse field components are adding. The~integrated field of such a magnet is an ideal dipole as the local non-dipolar components of the field in the ends are all self cancelling when integrating over the full axial length. Higher order, n-polar, fields can also be made with shapes that are not a smooth helix but have n "wiggles" per turn. In Figure ~\ref{fig:coilgeom} we can see illustrations of the three coil geometries for dipoles.

% SECTION
\section{Mechanical support and Pre-stress}
\begin{figure}[tb!]
\begin{center}
\includegraphics[height=4 cm]{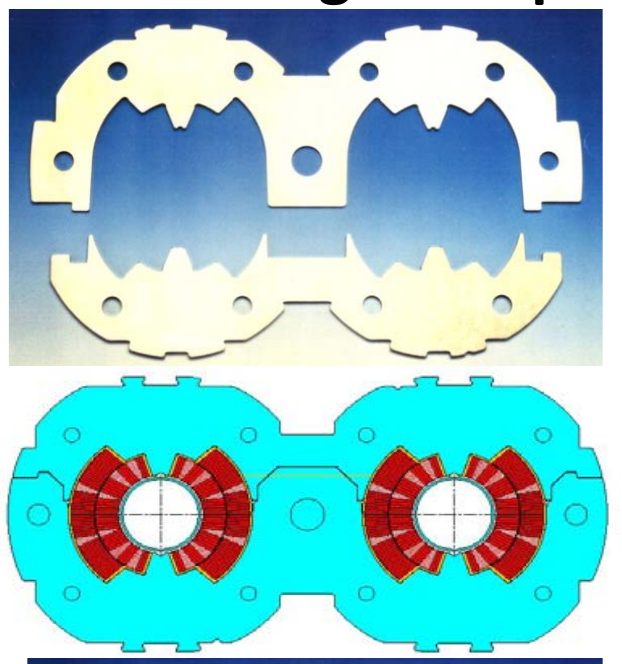} \hspace{0.5 cm}
\includegraphics[height=4 cm]{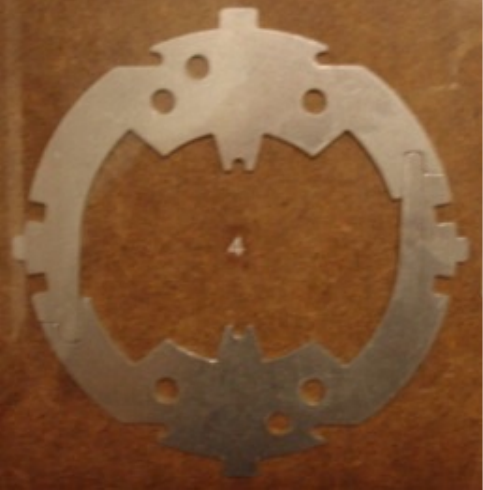}\hspace{0.5 cm}
\includegraphics[height=4 cm]{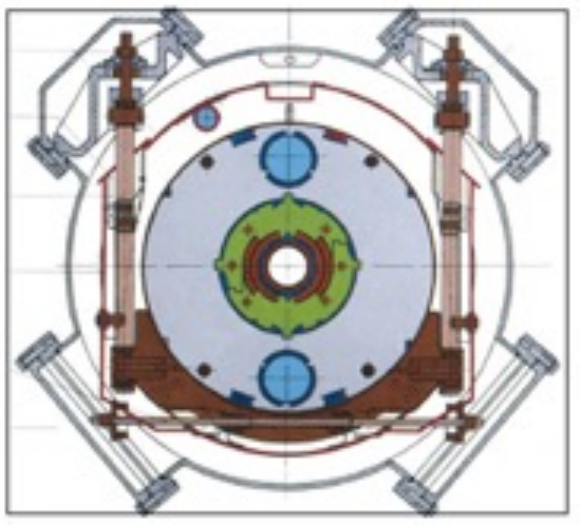} \hspace{1.5 cm}
\includegraphics[height=4 cm]{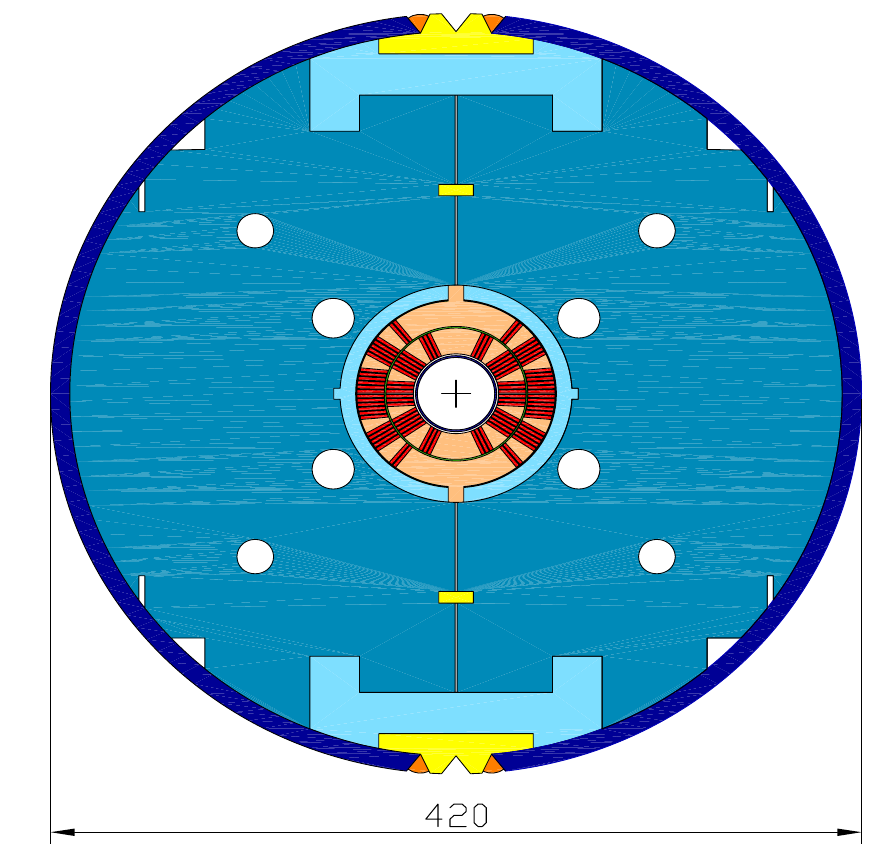} \hspace{1.5 cm}
\includegraphics[height=4 cm]{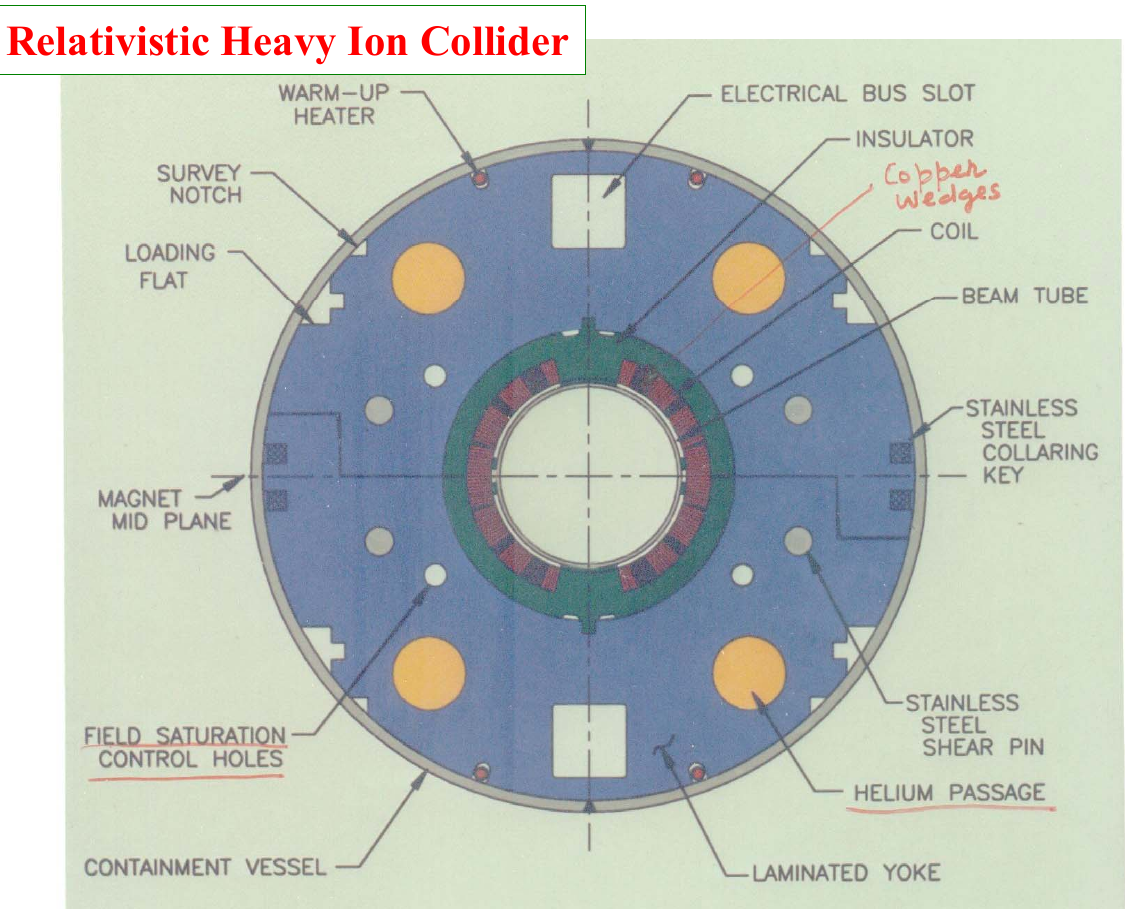} \hspace{1. cm}
\includegraphics[height=3.5 cm]{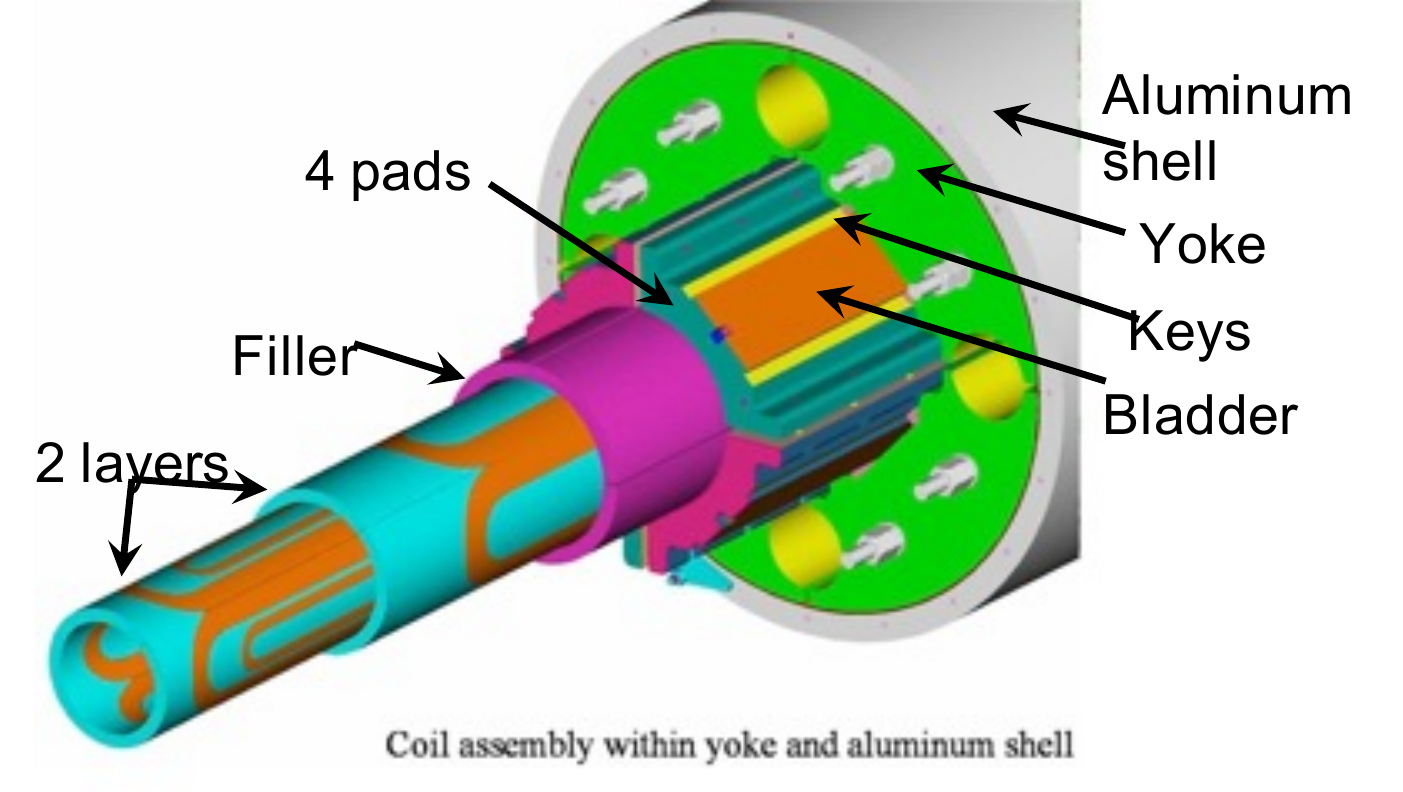} \hspace{0.3 cm}
\includegraphics[height=3.5 cm]{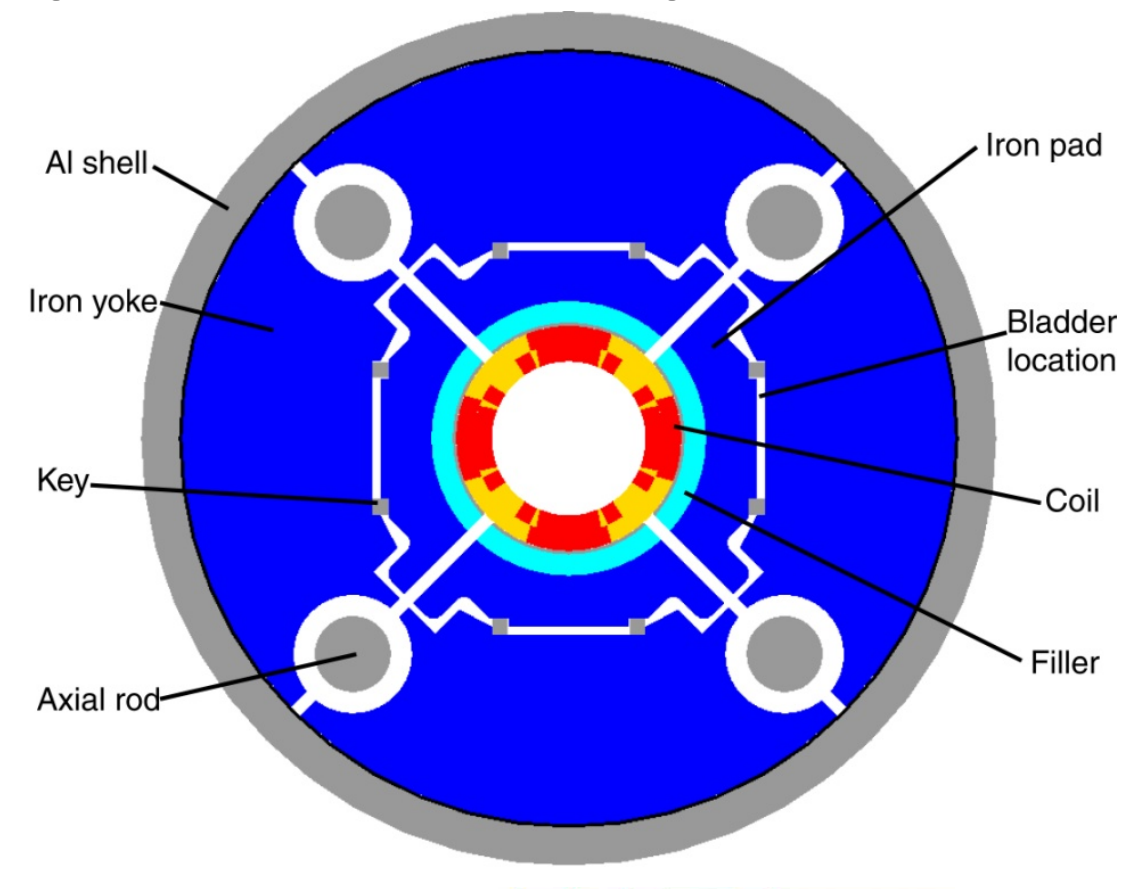} \hspace{0.3 cm}
\includegraphics[height=3.5 cm]{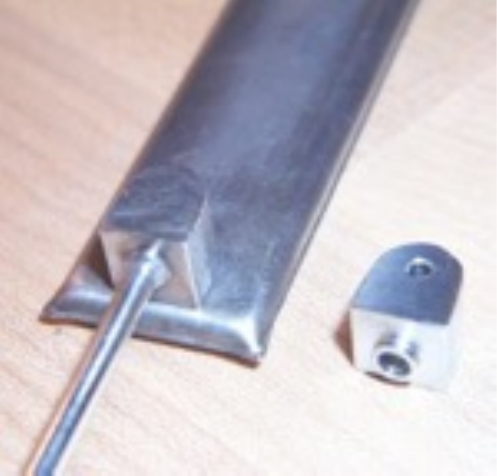} \hspace{0.3 cm}
\caption{Collars; left: LHC dipole; middle and right: HERA dipole (DESY). 
Middle row: Cylinder and keys; left: HFDA (FNAL); right: RHIC (BNL). 
Bottom row: Shell, bladder and keys; left and middle: TQS (LARP); right: Bladder.}
\label{fig:prestress}
\end{center}
\end{figure}

In superconductor accelerator magnets the field quality is of prime importance. To achieve the limits set on the field multipoles of only of few units, the positioning of some critical cables in the magnets has to be precise to $\sim0.02\Umm$. We have also seen that the coils are undergoing electromagnetic forces in the M$\UNZ$ range, that during the acceleration ramp of the accelerator also ramp up from a low value to these high values. Due to this the coils will inevitably both move and be compressed. These displacements can amount to several $\UmmZ$ and hence significantly deteriorate the field quality. At the same time a moving coil will generate heat, that with the low heat capacity of materials at low temperature will result in large temperature rises in the coil and can thus quench the coil. This can in some coils already happen at displacements of $\sim0.01\Umm$. The solution for these two problems is to put the coils under pre-stress to fix their position into the high field situation so that they practically don't move anymore under the electromagnetic forces. \\
We can look at two examples: the LHC dipole magnets with $B=8.34\UT$ that need a pre-stress of 30 M$\UPaZ$ and the Fresca2 dipole with $B=13\UT$ that needs a pre-stress of 130 M$\UPaZ$, both of course at cold temperature. 
For superconducting magnets there are roughly three commonly used methods to put pre-stress on coils:

\begin{itemize}
\item collars
\item shrinking cylinder and/or pre-stress key
\item Shell, bladder and keys.
\end{itemize}
All three methods will have to take into account the different shrinkage of the various components during cool-down and all the detailed geometrical and mechanical stress effects. Only very rough "ballpark" stress values can be calculated with analytical methods. For all superconducting magnets detailed finite element mechanical models are needed during the design and fabrication or order to get the required pre-stress everywhere and to prevent to over-stress the coil in parts or as a whole, at warm, during cool-down and when powering. Over-stress can lead to insulation failure or worse breaking the conductor. In the~next sections we will go a bit more in detail on these three methods (see  Fig. ~\ref{fig:prestress}).

% SUBSECTION
\subsection{Pre-stress: collars}
Providing pre-stress with collars is the classical solution that has been applied on most of the main magnets of accelerators up to now. Thin collars (laminations) are put around the coil that define a fixed cavity that is smaller that the coil at rest. The collars are compressed together in a press and locked with either keys or pins. The~collars can be made of an Al alloy or of a stainless steel. Depending on the material choice, due to the differential shrinkage during cool-down between the coil and the collars, the room temperature ($300\UK$) pre-stress can be a factor 2-3 times higher than the pre-stress at low temperature ($4.2\UK$). For the LHC dipoles, that have stainless steel collars, we have to put 70 M$\UPaZ$ at room temperature to get 30 M$\UPaZ$ at low temperature.\\

The fixed cavity at low temperature determines the shape of the coil and thus is a powerful tool to fix the field quality of the magnet. It should be clear that if the coil size or the collar shape is not well controlled the stress can end up either to low or too high so a strict control of the dimensions of all components is required. For high field Nb$_{3}$Sn magnets, the stress overrun at room temperature can be a~problem as with over-stress one can break the filaments in the wires.

% SUBSECTION
\subsection{Prestress: shrinking cylinder and/or prestress key}
One can provide pre-stress on coils using a split iron yoke that pushes on the coils. On the outside of the yoke a cylinder or a set of keys of a faster shrinking material presses the two half yokes together during cool-down. The pre-stress completely depends on the dimensioning of the components and the chosen materials. Also here, the shrinkage of all components during cool-down has to be taken into account.

% SUBSECTION
\subsection{Prestress: Shell, bladder and keys}
This method was developed at LBNL around 20 years ago. The coil is surrounded by a segmented steel yoke and a thick aluminium shell. At room temperature inflatable bladders are inserted in slits between the yoke and the coil pack and are pressurised with water (up to 600 bar) and fixed by inserting precision keys. Around half of the required pre-stress is applied at room temperature (between 10  M$\UPaZ$ and 80 M$\UPaZ$ ). When cooling down, the differential shrinkage between the Al shell and the yoke provides the rest of the pre-stress (up to a total of 150 M$\UPaZ$). This method never applies an over-stress and only at low temperatures the final pre-stress is present. The total pre-stress force and the EM forces are take by the~Al shell.  The whole process needs careful mechanical FE modelling before, and strain measurements, during bladder operations and cool-down.

% SECTION
\section{Operating superconducting magnets in accelerators}
Starting in the 1970s superconducting magnets have been employed in increasing numbers in accelerators. Initially, they were used in several beam-lines, e.g. at the SPS at CERN, principally to get experience with superconducting technology in this application. For this they were used at a fixed field or gradient and due to the single pass beam there were no specifically tight field quality demands. In the same era, the first low-beta quadrupoles were introduced in colliders, like the ISR. For this application the field quality was very important as field errors situated at high beta zones, like inside the low-beta quadrupole magnets, have a large effect on the beam. These colliders had very slow (many minutes) energy ramps and thus the dynamic effects on the beam were still limited. Superconducting cyclotrons have either one big central or several segmented magnets that still have iron poles to shape the field. These iron poles will dominate the field quality and attenuate the dynamic effects in to the superconducting coils. In synchrotrons with many, or with the majority superconducting magnets special care has to be taken to cope with the specific properties of these magnets, like:.
\begin{itemize}
\item ramp rate,
\item excitation curves and calibration,
\item persistent current decay and snapback,
\item machine parameter tuning and hysteresis effects,
\item cryogenic system operation,
\item continuous cryostats.
\end{itemize}
For keeping dynamic field effects under control thin filaments are needed in the wires (see Section 5). Magnet strings that are situated inside long cryostats have long warm-up and cool-down times, which render any repair very tedious. For this superconducting magnets with the entire cryogenic system have to be engineered to very high reliability standards. The relatively slow ramp-up and ramp-down of superconducting accelerators imply that it is not evident to do quick trails with the beam settings. To get to a good efficiency, careful preparation with appropriate computer simulations are needed before trying new setting parameters out on the beam.

% SECTION
\section{High field magnet development programs}
Accelerators and especially colliders for High Energy Physics are at the energy frontier and are using the~most advanced magnet technology available to reach the highest possible energy and luminosity. Upgrades for such machines and new machines will demand new types of magnets that exceed the field levels of the previous ones. To do this, extensive magnet development programs are needed that can take many years to yield the next generation of operational accelerator magnets. I will give a few examples of such magnet development programs that were running in the near past and presently.

% SUBSECTION
\subsection{Magnet development for HL-LHC}
\begin{figure}[ht!]
\begin{center}
\includegraphics[height=6 cm]{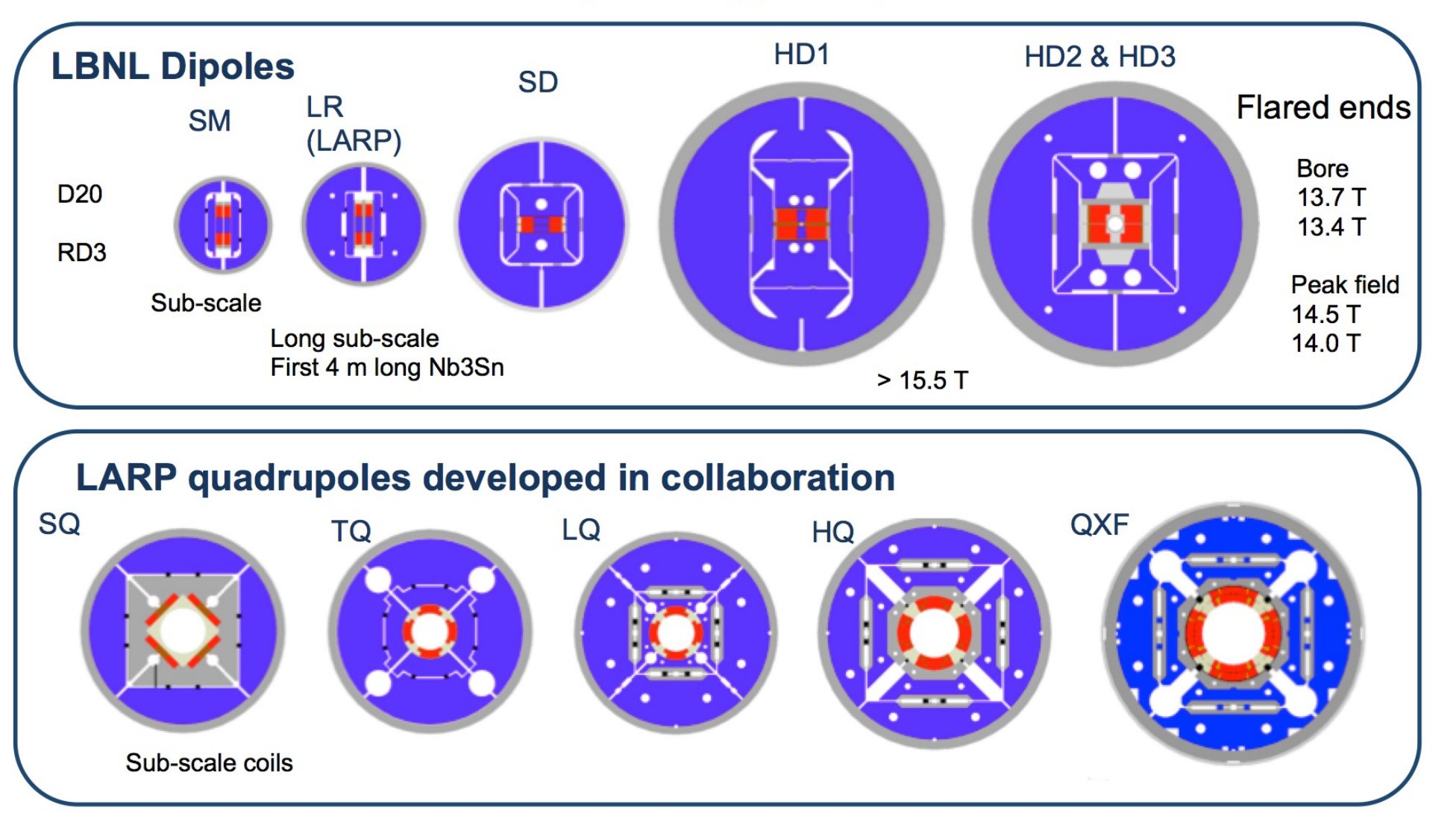}
\caption{Low beta Nb$_{3}$Sn quadrupole development for the HL-LHC: the US program (LARP) and the LBNL Nb$_{3}$Sn dipole development program.}
\label{fig:HILUMI-US}
\end{center}
\end{figure}

\begin{figure}[ht!]
\begin{center}
\includegraphics[height=11 cm]{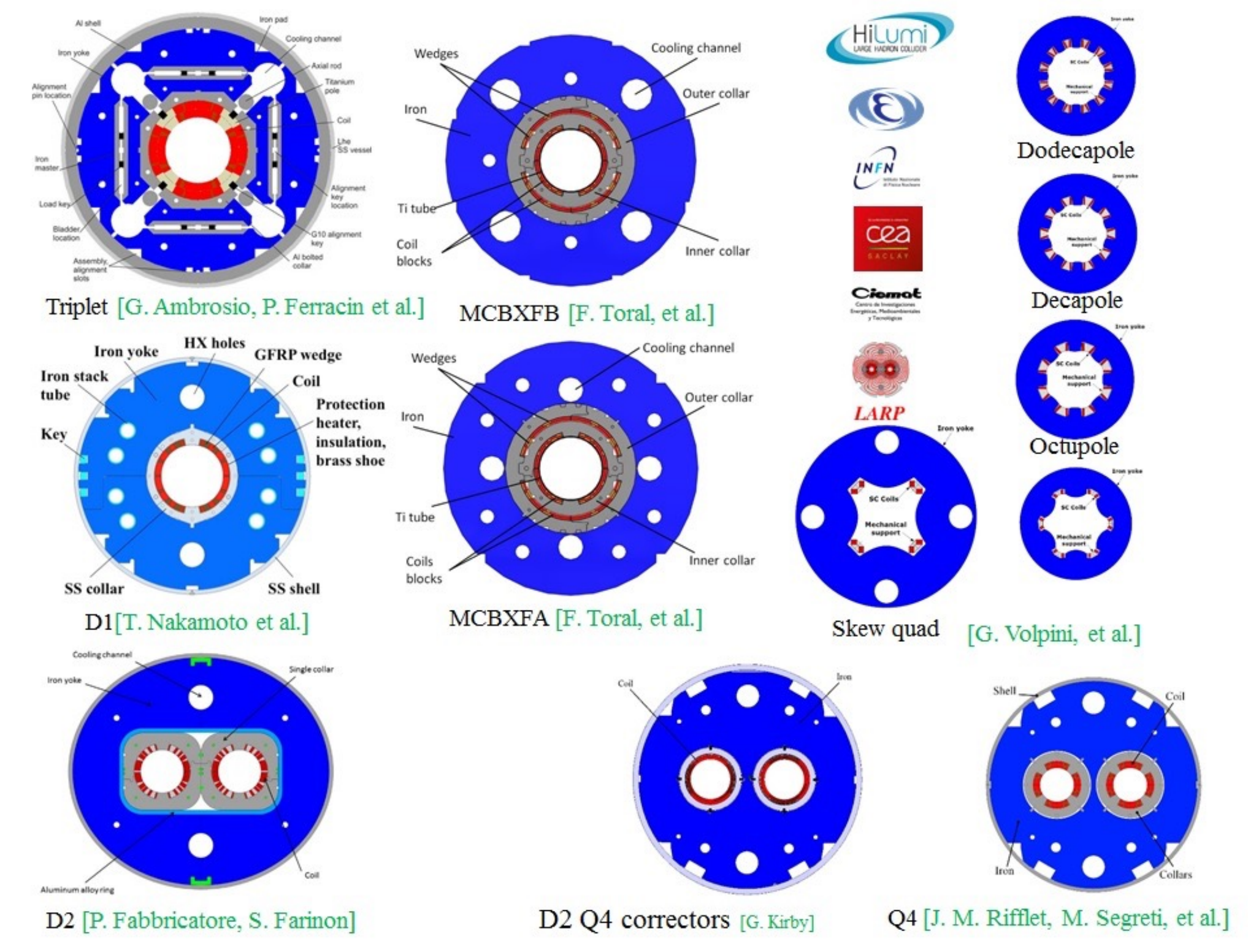}
\caption{HL-LHC low beta insertion magnet types.}
\label{fig:HILUMI-zoo}
\end{center}
\end{figure}

In Figure~\ref{fig:HILUMI-US} (bottom) we can see a schematic representation of the quadrupole development steps taken by the LARP collaboration in the US to develop the HL-LHC low beta insertion quadrupole MQXF, that has a gradient $G = 132 \UT / \UmZ$ in a $150 \Umm$ diameter aperture. In Figure~\ref{fig:HILUMI-zoo} we can see all the magnet types that are needed for the HL-LHC high luminosity insertions. More details on these insertion magnets can be found at https://espace.cern.ch/HiLumi/WP3/SitePages/Home.aspx.

% SUBSECTION
\subsection{Development of Nb$_{3}$Sn dipoles for future colliders.}
\begin{figure}[ht!]
\begin{center}
\includegraphics[height=9 cm]{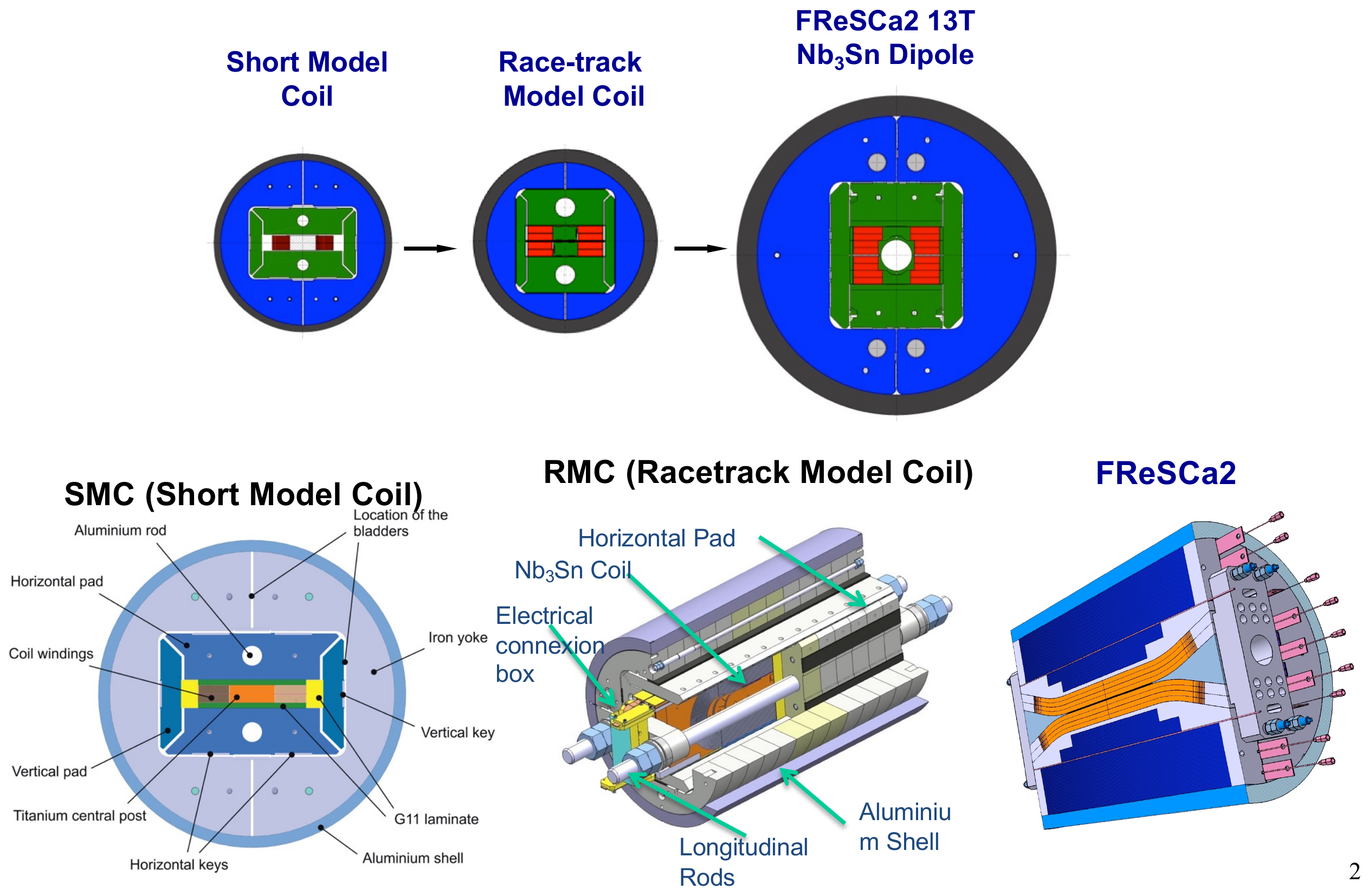}
\caption{CERN and CEA Nb$_{3}$Sn Fresca2 13 T dipole development program 2009-2017}
\label{fig:cern-cea}
\end{center}
\end{figure}

\begin{figure}[ht!]
\begin{center}
\includegraphics[height=5.5 cm]{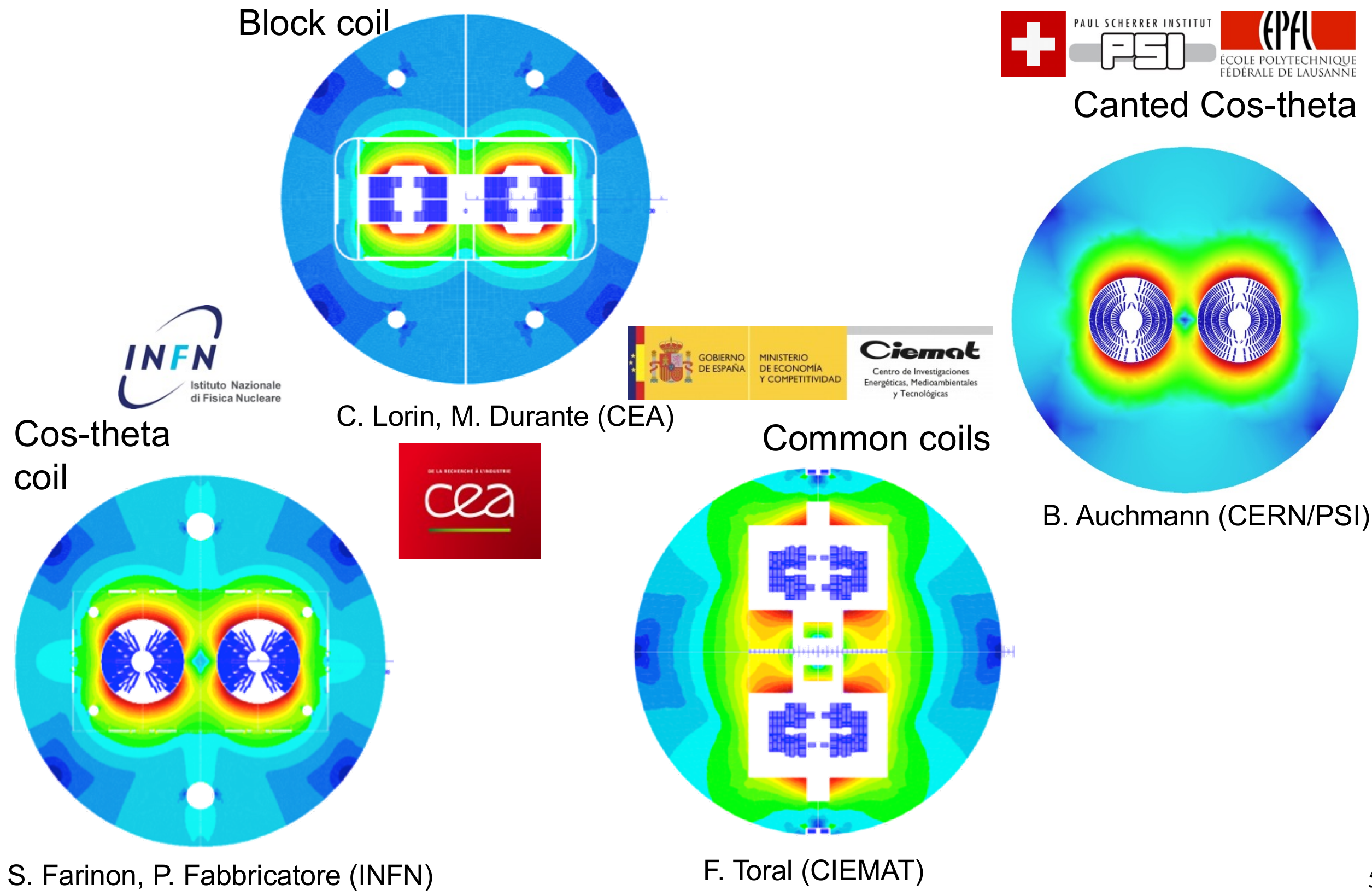} \hspace{0.5 cm}
\includegraphics[height=5.5 cm]{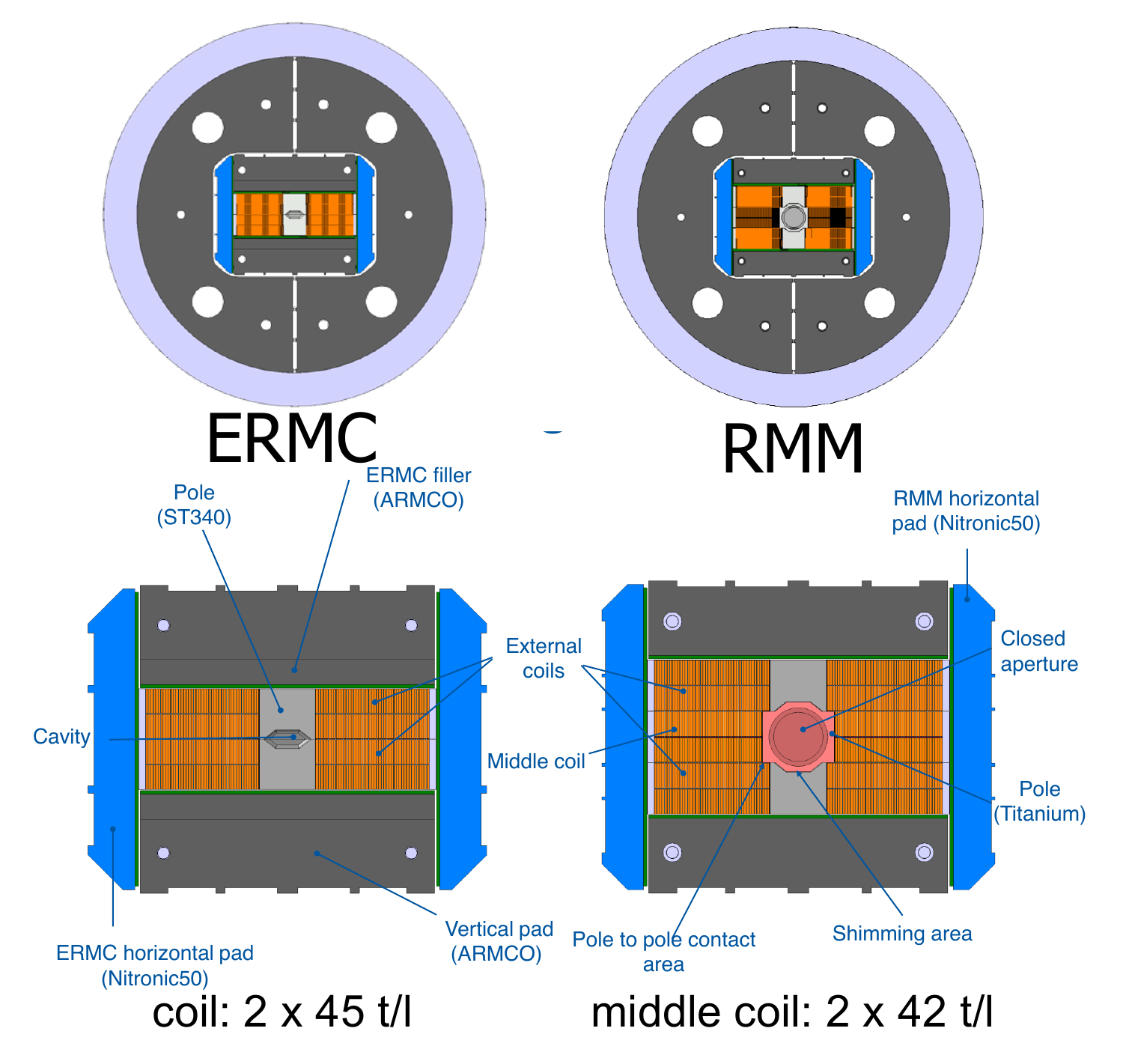}  
\caption{left: FCC 16 T dipole development program by European partners; right: First two steps of the CERN FCC 16 T Nb$_{3}$Sn dipole development}
\label{fig:fcc}
\end{center}
\end{figure}

In the first decade of this century LBNL in the US did some groundbreaking work with the development of shell-bladder-and-key structures for Nb$_{3}$Sn magnets. In the top picture of Fig.~\ref{fig:HILUMI-US} we can see the~development line for these dipoles that cumulated in $13.3\UT$ in the HD2 magnet.  At CERN and CEA, this technology was further developed in 2009-2017 with the stepwise development (see Fig.~\ref{fig:cern-cea}) of the~Fresca2 dipole magnet that reached a field of $14.6\UT$ in a $150\Umm$ diameter aperture. 
Since 2014 an international collaboration is working on the development of a $16 \UT$ dipole for the Future Circular Collider (FCC) that is being proposed at CERN. In Figure~\ref{fig:fcc} (left) we can find four types of dipoles that are under development in various European institutes. In parallel the US has its own development program for such magnets. At CERN, in continuation of the~Fresca2 magnet line, a stepwise development is underway with the eRMC and RMM models that will later be followed by a block coil magnet with flared end (see Fig.~\ref{fig:fcc} right) for which the first models are under test.

\newpage
% SUBSECTION
\subsection{Development of HTS dipoles for future colliders}
\begin{figure}[ht!]
\begin{center}
\includegraphics[height=5.5 cm]{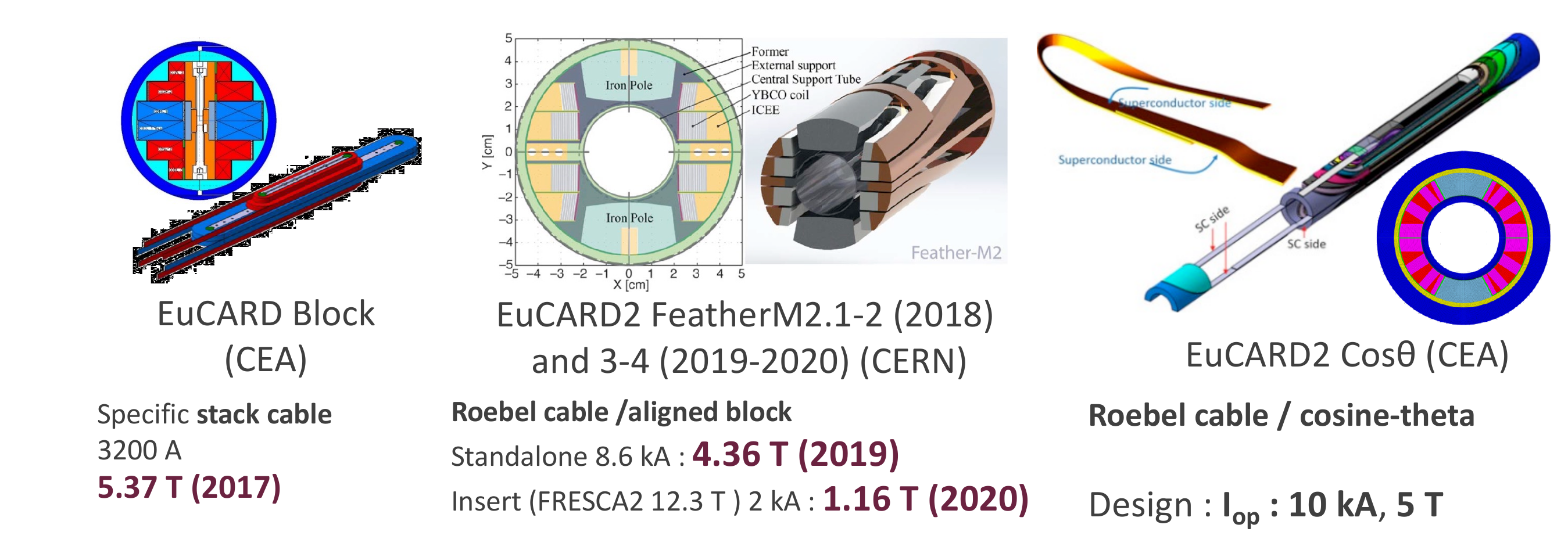}
\caption{First HTS dipoles made by CEA and CERN: left: EuCARD1 block coil using tape stacks (CEA); middle:~EuCARD2 block coil using Roebel cable (CERN); right: EuCARD2 $\cos\Theta$ using Roebel cable (CEA)}
\label{fig:hts-insert}
\end{center}
\end{figure}

To reach above $16 \UT$ HTS conductor will be needed. Practical HTS conductors are relatively recent and still being developed both in performance and in their industrial production technology resulting in constantly improving performance and slowly lowering prices. HTS accelerator magnets are really at the beginning of the development. In the last years in the US, Europe, China, Japan and South Korea efforts are underway to design and build the first small size magnets. In Europe, funded by the EU's EuCARD projects, three novel designs have been build and tested. In Figure ~\ref{fig:hts-insert} we can see these three designs that are using insulated cables with ReBCO tapes. In parallel, many institutes are   looking at coils without inter-turn insulation, or better said with controlled inter-turn resistance, that offer the potential to solve the transition (quench) problems that HTS magnets can experience. HTS magnets have large temperature margins and large temperature margins render magnets more robust to quenching but when a quench happens the propagation is very slow and thus very hard to detect and thus a local burnout can happen. Non-insulated coils offer alternative current paths around quench zones. The challenge in such coils is to gauge the resistances such that the ramp-up and ramp-down time constants of the field are acceptable for the application in an accelerator.

% SECTION
\section*{Acknowledgements}
For this lecture I used material from lectures, seminars and reports from many colleagues. Special thanks goes to: Giorgio Ambrosio (FNAL), Luca Bottura (CERN),  Shlomo Caspi (LBNL), Arnaud Devred (ITER and CERN), Paolo Ferracin (LBNL), Attilio Milanese (CERN), Jeroen van Nugteren, Juan-Carlos Perez (CERN), Lucio Rossi (CERN), Stephan Russenschuck (CERN), Ezio Todesco (CERN), Davide Tommasini (CERN) and Martin Wilson

% SECTION

%END
\end{document}